\documentclass[twocolumn]{aa}
\usepackage{natbib}
\usepackage{graphicx}
\usepackage{txfonts}

\newcommand{\myarcsec}{\hbox{$.\!\!^{\prime\prime}$}}

\newcommand{\myarcsecnodot}{\hbox{$\;\!\!^{\prime\prime}\;$}}

\newcommand{\angstrom}{\AA}

\def\cm3{\rm ~cm^{-3}}
\def\kms{\rm ~km~s^{-1}}

\def\ergsm{\rm ~erg~s^{-1} cm^{-2}}

\def\lsim{\!\!\!\phantom{\le}\smash{\buildrel{}\over
  {\lower2.5dd\hbox{$\buildrel{\lower2dd\hbox{$\displaystyle<$}}\over
                               \sim$}}}\,\,}
\def\gsim{\!\!\!\phantom{\ge}\smash{\buildrel{}\over
  {\lower2.5dd\hbox{$\buildrel{\lower2dd\hbox{$\displaystyle>$}}\over
                               \sim$}}}\,\,}
\begin{document}
\title{The outer rings of SN~1987A\thanks{Based on observations made with ESO Telescopes at the Paranal Observatory, Chile (ESO Programs 70.D-0379, and 082.D-0273A)}}
\titlerunning{The outer rings of SN~1987A}
\author{
A. Tziamtzis\inst{1}\and
P. Lundqvist\inst{1}\and
P. Gr{\"o}ningsson\inst{1}\and
S. Nasoudi-Shoar\inst{3}
          }
\offprints{A. Tziamtzis}

\institute{Stockholm Observatory, AlbaNova Science Center, Department of Astronomy, SE-106 91 Stockholm, Sweden
\email{anestis@astro.su.se}
\and
Argelander-Institut f\"ur Astronomie, Auf dem H\"ugel 71 D-53121 Bonn, Germany}
\date{Received 12/08/10 ; accepted 02/11/2010}
 
\abstract
{}
{We investigate the physical properties and structure of the outer rings of SN 1987A to understand their formation and evolution.}
{We used low resolution spectroscopy from VLT/FORS1 and high resolution spectra from VLT/UVES to estimate the physical conditions in the outer rings, using nebular analysis for emission lines such as [O~II], [O~III], [N~II], and [S~II]. We also measured the velocity at two positions of the outer rings to test a geometrical model for the rings. Additionally, we used data from the HST science archives to check the evolution of the outer rings of SN 1987A for a period that covers almost 11 years.}
{We measured the flux in four different regions, two for each outer ring. We chose regions away from the two bright neighbouring stars and as far as possible from the inner ring and created light curves for the emission lines of [O~III], H$\alpha$, and [N~II]. The light curves display a declining behaviour, which is consistent with the initial supernova-flash powering of the outer rings. The electron density of the emitting gas in the outer rings, as estimated by nebular analysis from the [O~II] and [S~II] lines, is $\lsim 3\times10^3$ cm$^{-3}$, has not changed over the last $\sim 15$ years, and the [N~II] temperature remains also fairly constant at $\sim 1.2\times10^4$ K. We find no obvious difference in density and temperature for the two outer rings. The highest density, as estimated from the decay of H$\alpha$, could be $\sim 5\times10^3$ cm$^{-3}$ however, and because the decay is somewhat faster in the southern outer ring than it is in the northern, the highest density in the outer rings may be found in the southern outer ring. For an assumed distance of 50 kpc to the supernova, the distance between the supernova and the closest parts of the outer rings could be as short as $\sim 1.7\times10^{18}$ cm. Interaction between the supernova ejecta and the outer rings could therefore start in less than $\sim 20$ years. We do not expect the outer rings to show the same optical display as the equatorial ring when this happens. Instead soft X-rays should provide a better way of observing the ejecta - outer rings interaction.}
{}

\authorrunning{A. Tziamtzis et al.}

\maketitle

\section{Introduction}

Owing to its proximity, SN 1987A in the Large Magellanic Cloud (LMC) has provided a unique opportunity for a detailed study of both pre-supernova mass loss and of the physical processes related to the supernova event itself. It is one of the few supernova explosions for which the progenitor star had been observed prior to the explosion. The triple ring system that surrounds the debris of the explosion presents a unique opportunity to check the models of the late stages of stellar evolution. The two outermost rings are roughly three times more extended than the inner ring (Wang \& Wampler 1997, Jakobsen et al. 1995) which is oriented north and south of the supernova, forming an hour-glass structure, and the most inner ring surrounds the supernova. A possible explanation for the formation of the rings is that they were formed by interaction of the wind from the progenitor star with gas that was released from the star at earlier stages of its evolution in the form of asymmetric wind-like structure (Blondin \& Lundqvist 1993, Chevalier \& Dwarkadas 1995).  

The ring system on SN 1987A was first detected through its $\lambda$5007 [O III] emission, 310 days after the explosion by Wampler \& Richichi (1989). The discovery led to further observations with ground-based telescopes (Crotts, Kunkel, \& McCarthy 1989, Wampler et al. 1990) and the HST (Jakobsen et al. 1991, Plait et al. 1995). The observations have shown that the inner ring of SN 1987A (henceforth referred to as the equatorial ring, ER) has an enhanced nitrogen abundance, which is N/C = 5.0 $\pm$ 2.0, and N/O = 1.1 $\pm$ 0.4 times the solar (Lundqvist \& Fransson 1996). These findings imply that the progenitor star was in a post He core burning phase at the time of the explosion (Podsiadlowski 1992). The ER has been studied extensively at optical wavelengths, and other wavelengths ranging from radio to X-rays. The optical observations are summarized and used in Mattila et al. (2010), who also gave an updated discussion of elemental abundances and gas densities of the ER. In particular, the He/H-ratio, by number, was found to be in the range 0.11 -- 0.23, which is lower than estimated by Lundqvist \& Fransson (1996), but still clearly higher than the solar value.

\begin{table*}
\caption{VLT/FORS1 Observational spectroscopic log.}
\begin{tabular}{c c c c c c c c}
\hline\hline
Source & Grism & $\lambda$ range & Exp. time  & Slit & Seeing & Airmass & Date\\
       &      &  (nm)&(s) &  (arcsec) & (arcsec)  &        \\
\hline
SN1987A & 600B & 345-590 & 1200  & 0.7 & 0.86 & 1.404 & 2002-12-30 \\
SN1987A & 600B & 345-590 & 1200  & 0.7 & 0.78 & 1.408 & 2002-12-30 \\
SN1987A & 600B & 345-590 & 1200  & 0.7 & 0.74 & 1.417 & 2002-12-30 \\
SN1987A & 600B & 345-590 & 1200  & 0.7 & 0.66 & 1.417 & 2002-12-30 \\
SN1987A & 600B & 345-590 & 1200  & 0.7 & 0.76 & 1.433 & 2002-12-30 \\
SN1987A & 600B & 345-590 & 1200  & 0.7 & 1.02 & 1.455 & 2002-12-30 \\
SN1987A & 600R & 525-745 &  960  & 0.7 & 0.72 & 1.418 & 2002-12-30 \\
SN1987A & 600R & 525-745 &  960  & 0.7 & 0.79 & 1.446 & 2002-12-30 \\
SN1987A & 600R & 525-745 &  300  & 0.7 & 0.87 & 1.448 & 2002-12-30 \\
SN1987A & 600R & 525-745 &  300  & 0.7 & 0.81 & 1.462 & 2002-12-30 \\
SN1987A & 600R & 525-745 &  960  & 0.7 & 0.72 & 1.496 & 2002-12-30 \\
SN1987A & 600R & 525-745 &  960  & 0.7 & 0.80 & 1.499 & 2002-12-30 \\
SN1987A & 600R & 525-745 &  960  & 0.7 & 1.02 & 1.563 & 2002-12-30 \\
\hline 
\hline
LTT3468 & 600B &  345-590 &100 & 0.8 & 0.68 && 2002-12-30 \\
LTT3468 & 600R & 525-745 &100  & 0.8 & 0.68 && 2002-12-30 \\ 
\hline
\end{tabular}
\end{table*}

The outer rings (henceforth ORs) have not been studied in the same detail as the ER because of their low surface brightness. Maran et al. (2000) found that the surface brightness of the ORs is only $\sim$ 5-15 \% of that of the ER. Ground based observations with NTT (Wampler et al. 1990) revealed the ORs in some detail, followed by HST observations (Burrows et al. 1995), which revealed their structure in even greater detail. The ORs extend $\sim 2\farcs5$ to the north, and 2\myarcsecnodot to the south of the supernova. Crotts et al. (1995) observed a structure that could connect the ORs with the ER. A feature like this can be real, but it can also be due to dust reflections, or light leakage from the ER and the supernova. Based on HST observations, Panagia et al. (1996) found that the ORs have lower CNO enrichment by a factor of three with respect to the ER. These findings were used as evidence to show that the ORs were formed $\sim$ 10,000 years before the ER. Kinematic observations by Crotts \& Heathcote (2000) disputed these findings and argued that all three rings were created $\sim$ 20,000 yrs before the supernova explosion. Finally, Maran et al. (2000) used HST/STIS to study both the ER and ORs, and found that the ORs have probably the same abundances as the inner ring. We will below use NOR and SOR for the northern outer ring and southern outer ring, respectively.

We decided to investigate the conditions in the ORs of SN 1987A in a more systematic way. To do this we used both spectroscopic (low-resolution spectra from FORS1 and high-resolution spectra from UVES on ESO-VLT), and archive photometric data from the HST (from WFPC 2 and ACS). We were able to measure the strength of the emission lines, checked their evolution over the time window between the two spectroscopic epochs ($\sim$ seven years), and used the photometry for checking the validitity of the flux measurements from the spectroscopic data. We also created light curves to study the evolution of the ORs based on the flux evolution of the [O III], [N II] and H$\alpha$ emission. Moreover monitored the evolution of the ratios of the [O III] and [N II] emission over H$\alpha$. In Sect. 2 we present the observations and the data reduction processes. In Sects. 3 and 4 we present the results from the spectroscopy and photometry, respectively. Finally, a summary and a discussion of the results is presented in Sect. 5.

\section{Observations}

\subsection{Spectroscopy with FORS 1}

Service mode observations of SN 1987A were done on 2002 December 30 with the FOcal Reducer/low dispersion Spectrograph 1 (FORS1) at ESO/VLT at Paranal, Chile. The detector has a scale of 0\myarcsec2 per pixel, and two different grisms were used together with a 0\myarcsec7 wide slit. The first grism covers the waverange between 3450-5900 \angstrom{} (ESO code GRIS600B+12), while the second covers the waveband between 5250-7450 \angstrom{} (ESO code GRIS600R+14). All spectra were obtained at the same slit position. To avoid effects from nearby stars the slit was centred to the SN and rotated to PA = 30$^\circ$. To minimize the effect of cosmic rays and bad pixels, the images were dithered with respect to each other, ensuring that any part of the supernova remnant was integrated over several pixels. More information about the observations can be found in Table 1.

\begin{figure*}
\centering
\includegraphics[width=16cm]{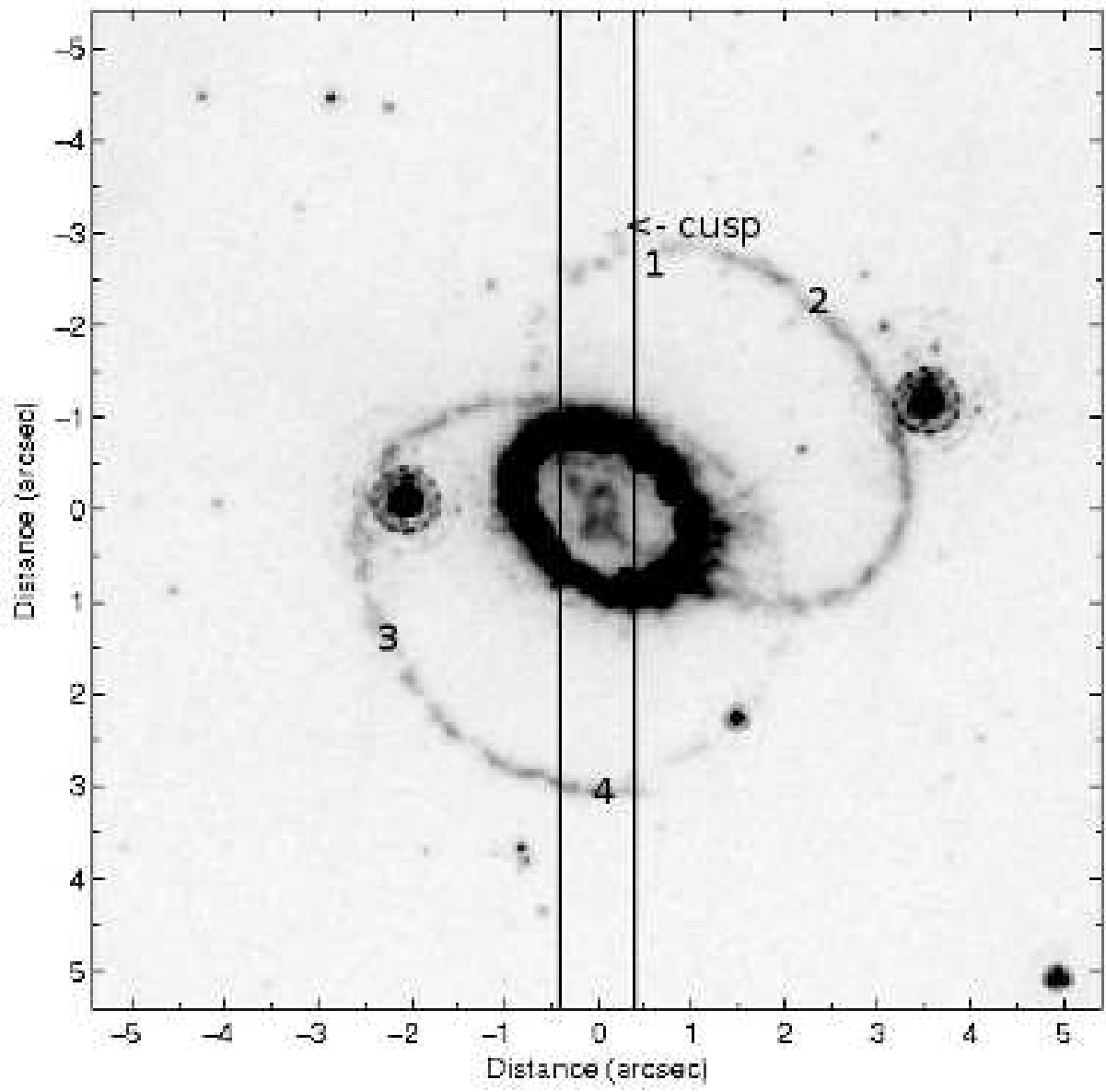}
\caption{HST/ACS image (filter F658N) of the triple ring system of SN~1987A from 2003 January (obtained by the SAINTS team; PI: R.P. Kirshner). The slit position of the VLT/UVES observations is superposed (the slit width is 0\myarcsec8 and PA=30$^\circ$). The numbers (1-4) correspond to the locations where we did the photometric measurements. Note how the NOR shows some irregularity with a cusp-like feature within the slit.}
\end{figure*}

\begin{table*}
\centering
\caption{VLT/UVES observations of SN 1987A and its rings.}
\begin{tabular}{c c c c c c c c c}
\hline
\hline
Date & Days after & Setting &$\lambda$ range &Slitwidth &Resolution &Exposure &Seeing &Airmass\\ 
& explosion&&(nm) &(arcsec) &$(\lambda/\Delta\lambda$) &(s) &(arcsec)\\
\hline
2002-10-4--7  & 5702--5705  & 346+580 & 303--388  & 0.8 & 50,000 & 10,200 & 0.7--1.0 & 1.5--1.6\\
2002-10-4--7  &             & 346+580 & 476--684  & 0.8 & 50,000 & 10,200 & 0.7--1.0 & 1.5--1.6\\
2002-10-4--7  &             & 437+860 & 373--499  & 0.8 & 50,000 & 9,360  & 0.4--1.1 & 1.4--1.5\\
2002-10-4--7  &             & 437+860 & 660--1060 & 0.8 & 50,000 & 9,360  & 0.4--1.1 & 1.4--1.5\\
2008-11-23    & 7944--8021  & 346+580 & 303--388  & 0.8 & 50,000 & 9000   & 0.8--1.0 & 1.4--1.5\\
2009-01-9--25 &             & 346+580 & 476--684  & 0.8 & 50,000 & 9000   & 0.8--1.0 & 1.4--1.5\\
2009-02-7--8  &             & 437+860 & 373--499  & 0.8 & 50,000 & 11,250 & 0.8--1.3 & 1.4--1.5\\
2009-02-7--8  &             & 437+860 & 660--1060 & 0.8 & 50,000 & 11,250 & 0.8--1.3 & 1.4--1.5\\
\hline
\hline
\end{tabular}
\end{table*}
\subsection{Spectroscopy with UVES}

High resolution spectra of SN 1987A were obtained with the Ultraviolet and Visual Echelle Spectrograph (UVES) at ESO/VLT at Paranal, Chile (Dekker et al. 2000). The UVES spectrograph disperses the light beam into two separate arms that cover different wavelength ranges of the spectrum. The blue arm covering the shorter wavelengths ($\sim3030-4990$ \AA) is equipped with a single CCD detector with a spatial resolution of 0\myarcsec246 per pixel, while the red arm ($\sim4760-10\,600$ \AA) is covered by a mosaic of two CCDs with a resolution of 0\myarcsec182 per pixel. Thus with two different dichroic settings the wavelength coverage is $\sim3030-10\,600$ \AA. Because of the CCD mosaic in the red arm there are, however, gaps at $5770-5830$ \AA{} and $8540-8650$ \AA. The spectral resolving power is $50\,000$ for a 0\myarcsec8 wide slit.

Spectra of SN 1987A were obtained at 2 different epochs, in 2002 October and in 2009 February, henceforth referred to as epochs 1 and 2 (see Table 2 for the observational details). The observations were performed in service mode for both epochs. A 0\myarcsec8 wide slit was centred on the SN and put at a position angle PA$=30^{\circ}$ (see Fig. 1) for both epochs. Hence, the spectral resolution for these epochs was $\lambda/\Delta\lambda\sim50\,000$, which corresponds to a velocity of $6\kms$.

\subsection{Optical data from HST}

Apart from the spectra, we downloaded photometric data from the HST science archive to study the evolution of the ORs. The images we acquired were taken between 1994 and 2005 with narrow band filters. We downloaded images from the Wide Field Planetary Camera 2 (WFPC 2), taken with the Planetary Camera (PC 1), which has a field of view of 35 $\times$ 35 arcseconds, and a scale of 0\myarcsec046 per pixel. The images from PC 1 were taken using the F502N [O III] filter ($\lambda_c$ = 5012 \AA, $\Delta\lambda$ = 27 \AA), the F656N H$\alpha$ filter ($\lambda_c$ = 6564 \AA, $\Delta\lambda$ = 22 \AA), and the F658N [N II] filter ($\lambda_c$ = 6591 \AA, $\Delta\lambda$ = 29 \AA).

We also used images taken with the Advanced Camera for Surveys (ACS) on the High Resolution Channel (HRC). The ACS on the HRC channel offers a field of view of 26 $\times$ 29 square arcseconds, and has a scale of 0\myarcsec025 per pixel. The ACS observations were done with the F502N [O III] ($\lambda_c$ = 5022 \AA, $\Delta\lambda$ = 57 \AA), and the F658N H$\alpha$ ($\lambda_c$ = 6584 \AA, $\Delta\lambda$ = 73 \AA) filters. In order to better understand the evolution of the outer rings, the data were selected to cover the longest possible time interval. Photometric observations that were made around the epoch of our spectroscopic observations were used also for checking the validity of the fluxes that we measured with FORS1 and UVES. In Table 3 a summary of the HST observations is presented.

\begin{table}
\caption{Observational photometric log of HST data.}
\begin{tabular}{c c c c c}
\hline\hline
Instrument & Filter  & Exp. time     & Date & Days after \\
       &      & (seconds) &        & the explosion\\
\hline
WFPC2 & F502N & 2400 & 1994-02-03 & 2524\\
WFPC2 & F502N & 4800 & 1994-09-24 & 2779\\
WFPC2 & F502N & 7800 & 1996-02-06 & 3281\\
WFPC2 & F502N & 8200 & 1997-07-12 & 3802\\
WFPC2 & F502N & 3600 & 2000-06-16 & 4872\\
WFPC2 & F502N & 5600 & 2000-11-14 & 5017\\
WFPC2 & F502N & 4200 & 2001-12-07 & 5410\\
ACS   & F502N & 8000 & 2003-01-05 & 5816\\
ACS   & F502N & 4000 & 2003-11-28 & 6130\\
ACS   & F502N & 3600 & 2004-12-15 & 6512\\
WFPC2 & F656N & 2400 & 1994-02-03 & 2524\\
WFPC2 & F656N & 5600 & 1997-07-10 & 3800\\
WFPC2 & F656N & 3100 & 1998-02-05 & 4011\\
WFPC2 & F656N & 7200 & 1999-01-08 & 4348\\
WFPC2 & F656N & 3800 & 1999-04-21 & 4452\\
WFPC2 & F656N & 2700 & 2000-06-16 & 4872\\
WFPC2 & F656N & 3200 & 2000-02-02 & 4738\\
WFPC2 & F656N & 4800 & 2000-11-14 & 5017\\ 
WFPC2 & F656N & 2200 & 2001-03-23 & 5154\\ 
WFPC2 & F656N & 4000 & 2001-12-07 & 5410\\ 
WFPC2 & F656N & 1500 & 2002-05-10 & 5567\\ 
WFPC2 & F658N & 4800 & 1994-09-24 & 2779\\ 
WFPC2 & F658N & 5200 & 1996-02-06 & 3281\\ 
WFPC2 & F658N & 5400 & 1996-09-01 & 3486\\
WFPC2 & F658N & 2300 & 2002-05-10 & 5567\\
ACS   & F658N & 8000 & 2003-01-05 & 5816\\
ACS   & F658N & 2800 & 2003-08-12 & 6026\\
ACS   & F658N & 4000 & 2003-11-28 & 6130\\
ACS   & F658N & 3560 & 2004-12-15 & 6512\\
ACS   & F658N & 2600 & 2005-04-02 & 6632\\
ACS   & F658N & 3600 & 2005-11-19 & 6852\\
\hline
\end{tabular}
\end{table}

\section{Data reduction}

\subsection{FORS1}

Apart from the science data, calibration data were obtained as well. These included bias and flat-field frames, along with spectra of calibration lamps and standard stars. The data reduction process was done by following standard procedures within IRAF. Initially the data were bias-subtracted and flat-field corrected by their corresponding master-bias and master-flat frame. To perform wavelength calibration, combined spectra of HeAr and HgCdHeAr arclamps were used. The line identification on the reference spectra was obtained within IRAF's task IDENTIFY, and the task FITCOORDS was later used for correcting possible distortion in our images. We checked the accuracy of our wavelength calibration with the strong skylines that are present in the science frames and the error is $\sim$ 0.1 \AA{} for the H$\alpha$ line. After the linear correction in the individual frames, the next step was to remove the background effects from the LMC and the sky lines. Background subtraction was done with IRAF's task BACKGROUND. A third-order Legendre polynomial was fitted in to estimate the background effects. Apart from a few lines where the LMC background is very strong (i.e H$\alpha$ and [O~III] lines), most of the LMC and sky effects are removed, but any effect from the remain is minimal. 

The spectra from the ORs of SN 1987A were extracted with IRAF's APALL task. The size of the aperture was chosen in each frame after careful examination to make sure that the contribution from the ER is minimal (see Fig. 1). As seen in Fig. 1 the signal-to-noise for the NOR is much weaker, and no spectra were extracted from there. The analysis from the FORS1 data is thus based on the spectra that we extracted from the SOR. Finally all 1-D images were combined by using a cosmic-ray rejection algorithm, and a spectrum of SN 1987A was obtained (see Fig. 4). We calibrated the flux with a spectrum of the spectro-photometric standard star LTT3864, which was observed in conjunction with SN 1987A. For the extinction we adopted a reddening of E$_{B-V}$ = 0.16 (Fitzpatrick \& Walborn 1990), with E$_{B-V}$ = 0.06 mag for the Milky Way (Staveley-Smith et al. 2003), and E$_{B-V}$ = 0.10 mag for the LMC. The reddening law was taken from Cardelli et al. (1989) using R$_V$ = 3.1.

\subsection{UVES}

For the data reduction we made use of the MIDAS implementation of the UVES pipeline version 2.0 for epoch 1, and version 2.2 for epoch 2. For a detailed description of the steps involved in the reductions we refer the reader to Gr\"oningsson et al. (2008a). For the time evolution of the fluxes an accurate absolute flux calibration is necessary. As discussed in Gr\"oningsson et al. (2006) and Gr\"oningsson et al. (2008a), we estimated the accuracy of the flux calibration by comparing the spectra of flux-calibrated spectro-photometric standard stars with their tabulated physical fluxes. In addition, we compared emission line fluxes from different exposures (and hence with different atmospheric seeing conditions) within the same epochs. From these measurements and the results from HST photometry presented in Gr\"oningsson et al. (2008b) we conclude that the accuracy in relative fluxes should be within $10-15\%$. The estimate of the absolute systematic flux error was done from a comparison between the UVES fluxes and HST spectra and photometry (see Gr\"oningsson et al. (2008ab). From this we find that the uncertainty of the absolute fluxes should be less than $20-30\%$. The resulting 1D spectra were average-combined, and cosmic rays were removed.

\begin{figure}[ht]
\centering
\includegraphics[width=9cm]{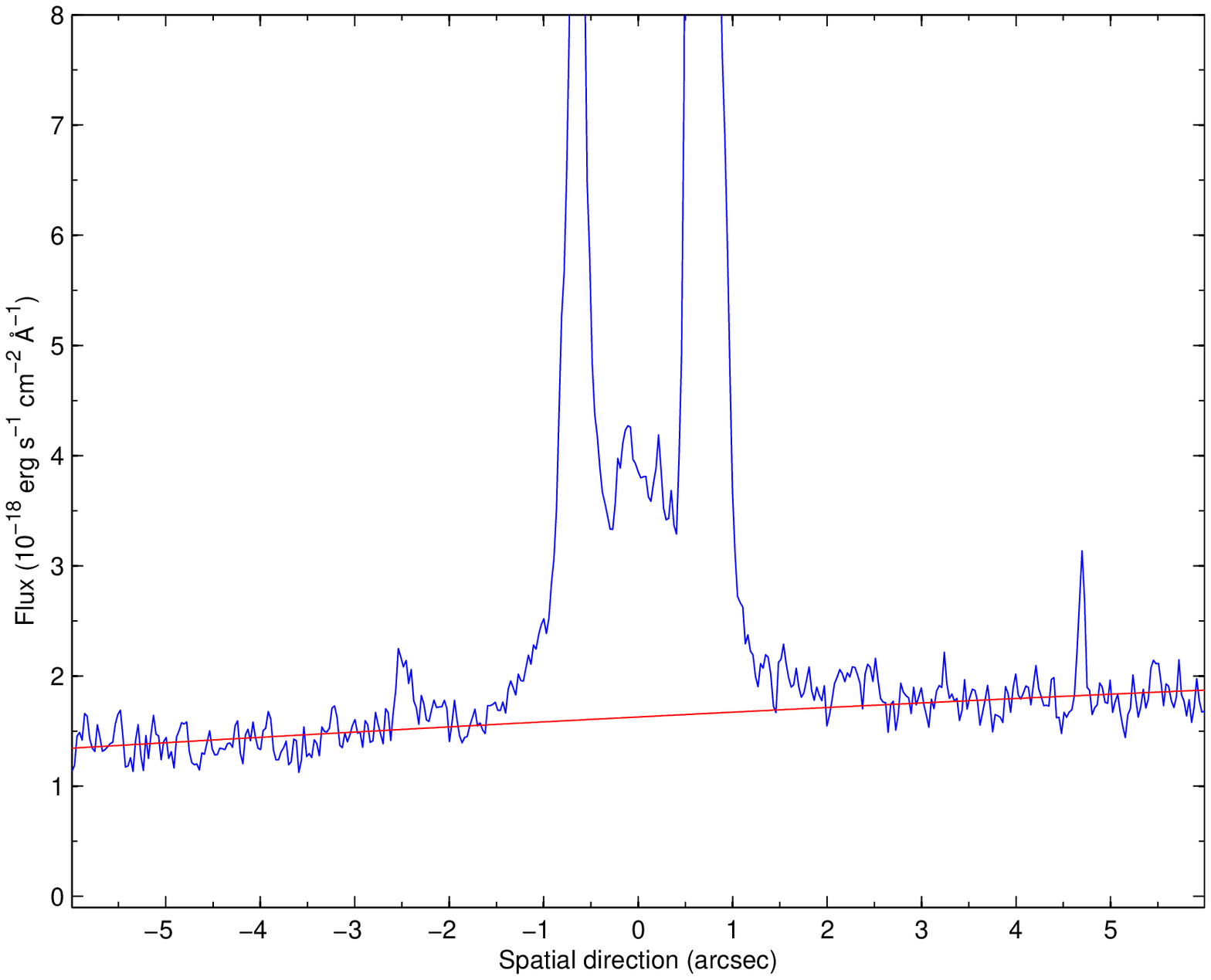}
\includegraphics[width=9cm]{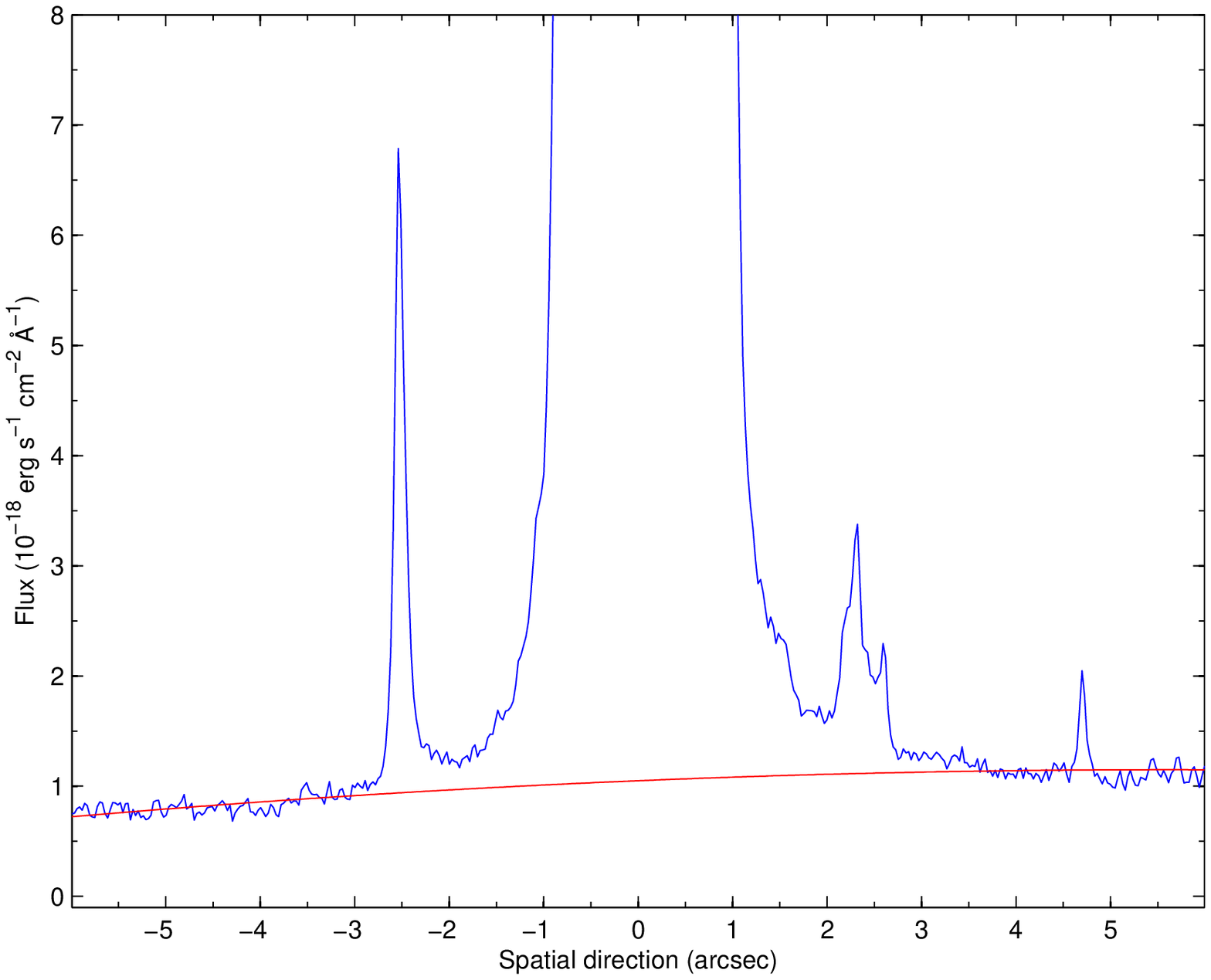}
\caption{{\it Top:} Spatial profile of the emission along the UVES slit (see Fig. 1) for [O~III] ~$\lambda$5007, created from the F502N filter of ACS. {\it Bottom:} Combined spatial profile of the H$\alpha$ and the [N~II]~$\lambda$$\lambda$ 6548, 6583 doublet created from the F658N filter of ACS (2003 January 5). In both plots the SOR is shown to the left and the NOR to the right. The cusp-like feature discussed in Fig. 1 is clearly seen as a broader outer-ring feature from the NOR than from the SOR. The spike at $\sim 4\farcs7$ to the north is a star.}
\end{figure}

\begin{figure}[ht]
\includegraphics[width=9cm]{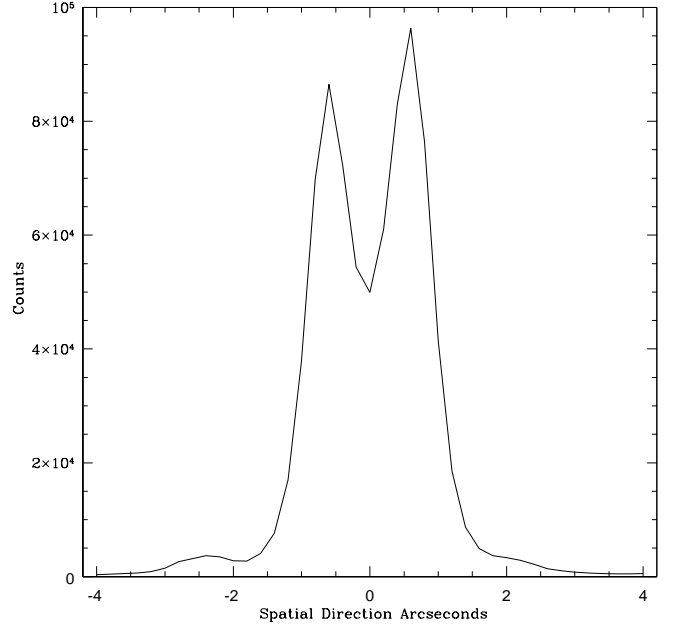}
\caption{Spatial profile of [N~II]~$\lambda$6583 across the slit. The SOR can be clearly seen on the left side of the figure at 2\myarcsec4 from the centre of the remnant. Because of seeing and resolution limitations the NOR on the right is not clearly visible. The image was made with the spectrum from FORS1 (2002 December 30).}
\end{figure}

\begin{figure}[ht]
\includegraphics[width=9cm]{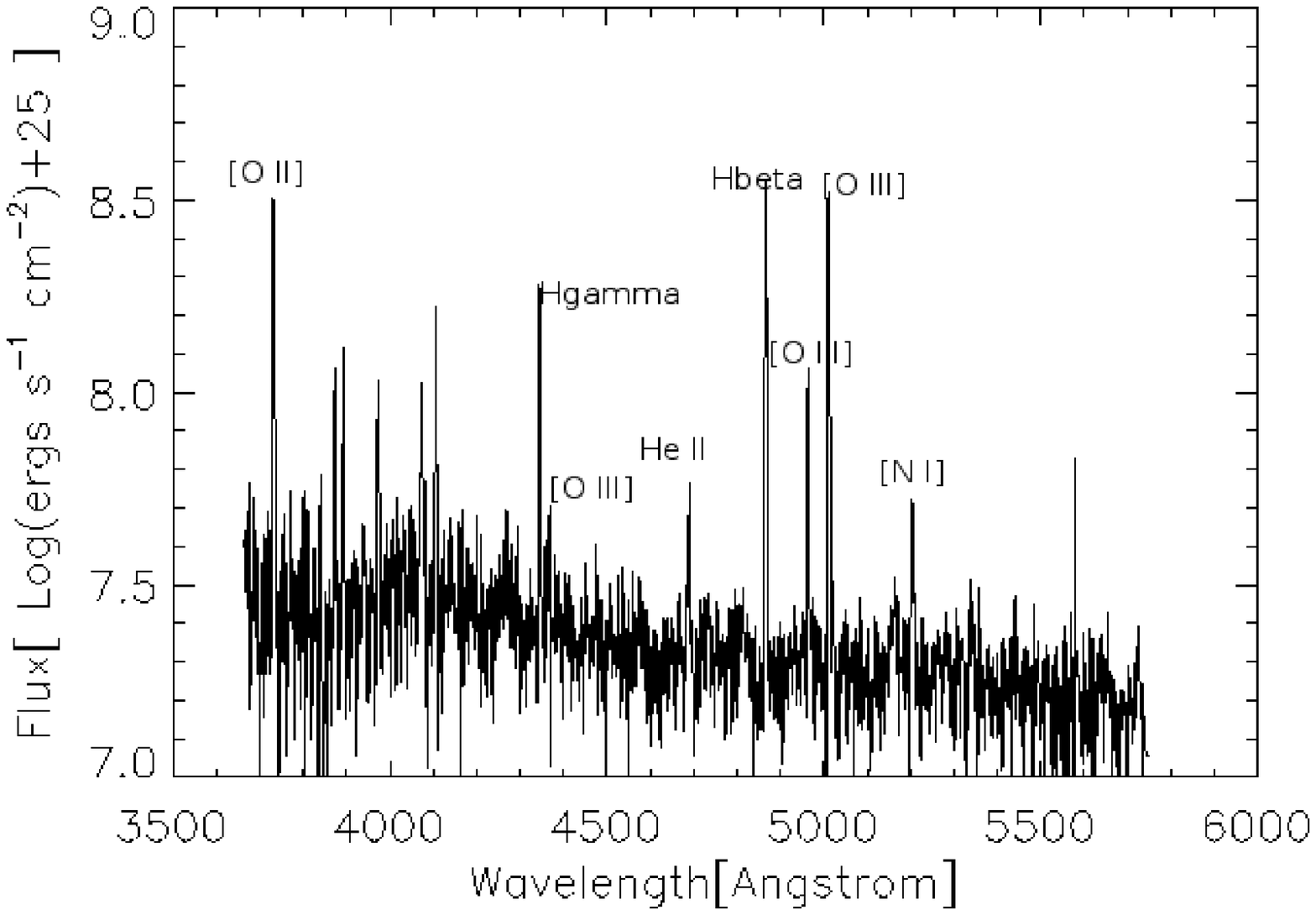}
\includegraphics[width=9cm]{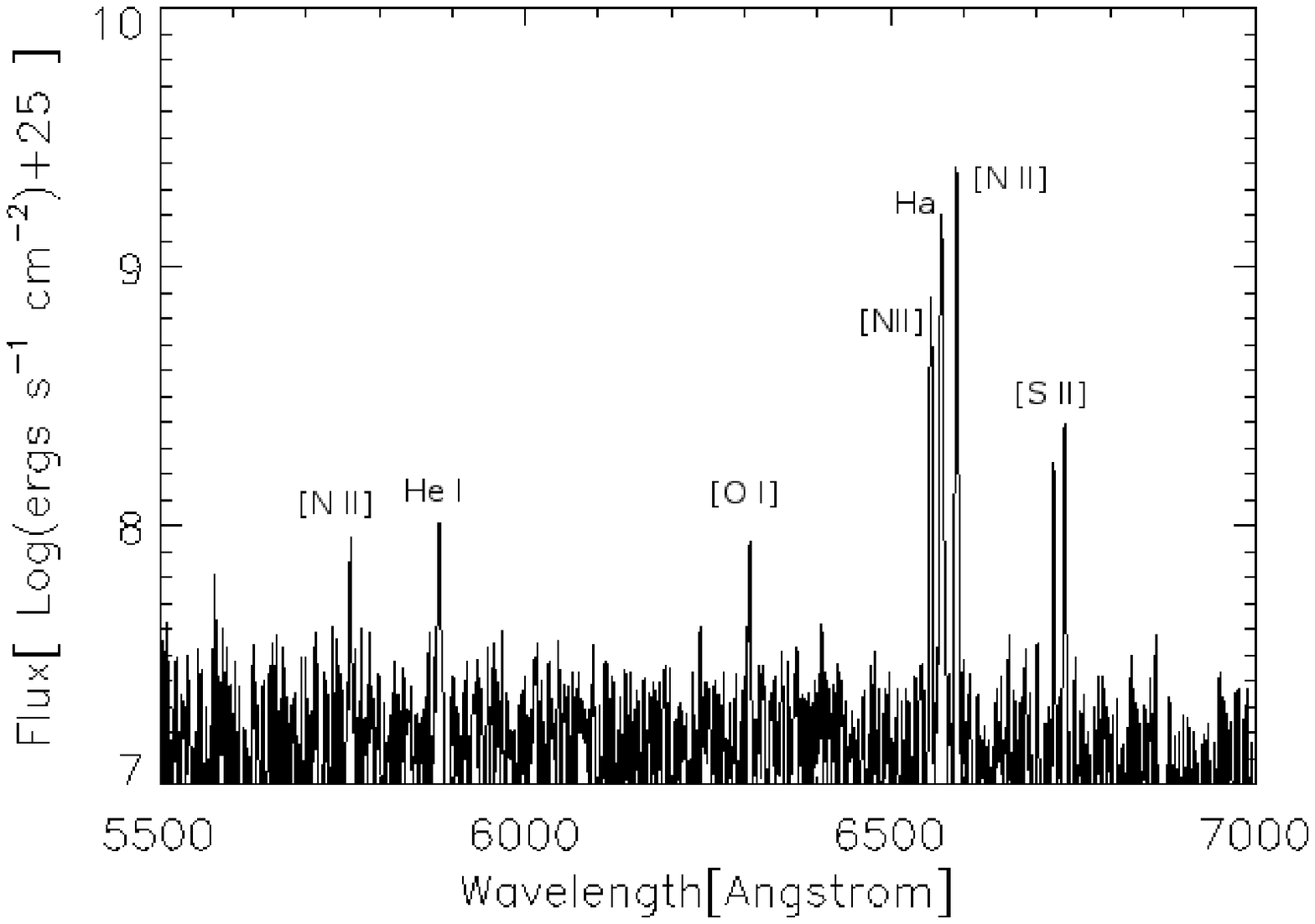}
\caption{{\it Top:} Spectrum of the SOR of SN 1987A (blue grism). {\it Bottom:} Spectrum of the SOR of SN1987A (red grism). Both spectra were taken with FORS1 on VLT on 2002 December 30.}
\end{figure}

\subsection{HST data}

For the HST data, we followed the standard procedures in IRAF for correcting our frames from the geometric distortions that are present. In this way the images from each epoch were sky-subtracted, corrected for the distortion, aligned, cleaned from cosmic rays, and drizzled into a single output count per second image. Additional masking was applied to remove the remaining bad pixels. These were identified manually and replaced by values of the neighbouring pixels through linear interpolation, using IRAF's task FIXPIX. This method can introduce uncertainties in the regions where the outer rings are present, but because of the high resolution of the HST frames this effect is low. The final images were aligned and scaled to the same size and image coordinates with the tasks GEOMAP and GEOTRAN. The absolute photometric flux calibration was achieved with the information given by the PHOTFLAM in the headers of the individual frames.

\begin{figure}[ht]
\includegraphics[width=8.5cm]{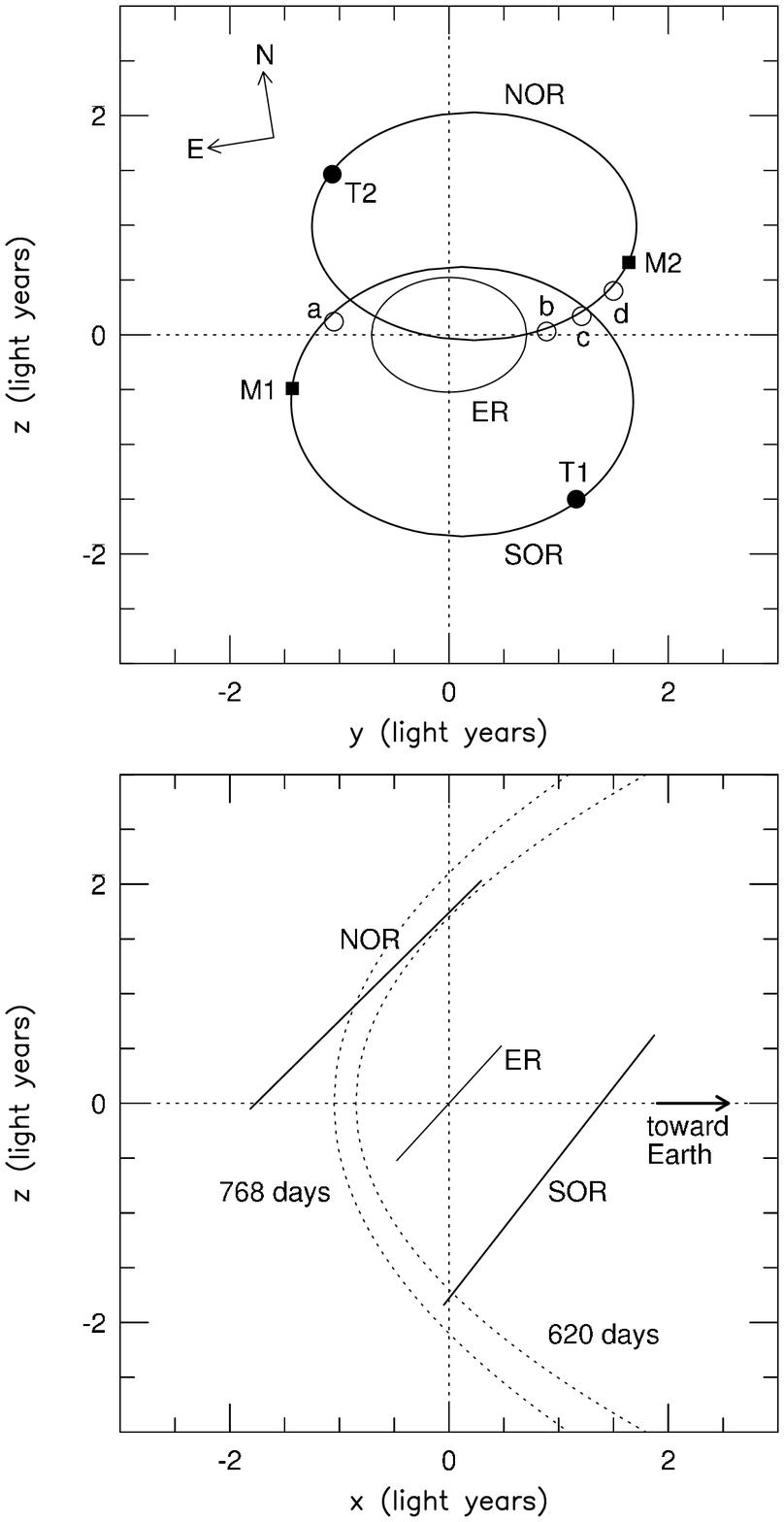}
\caption{{\it Top:} Model for the ER and ORs assuming tilted circular rings. The position of the rings was fixed to the observations of Crotts \& Heathcote (2000) (see text). {\it Bottom:} The tilted rings seen from the side. Light-echo paraboloids for 620 and 768 days after the explosion are shown. At those epochs the light-echo paraboloids intersect the ORs at the positions (named T1 and T2 in the upper panel, and named areas 4 and 1 in Table 9) observed through the FORS 1 and UVES slits.}
\end{figure}

\section{The geometry and physical conditions of the outer rings}

\subsection{Geometry}

We first made an attempt to provide the 3D geometry of the triple ring system. As seen in Figs. 1 and 5, the points T1 and T2 in Fig. 5 are the areas within the spectroscopic slit used for the UVES and FORS1 observations. The emission from these areas (called areas 4 and 1 respectively in Fig. 1 and Table 9) arrive at Earth at different times, and are delayed with respect to the supernova breakout by different amount of time. This light-echo effect for the ER was discussed in Lundqvist \& Fransson (1996), assuming a tilted ring centred on the supernova. The ORs are displaced from the equatorial plane, and may not be tilted in the same way as the ER. Both these effects are important for the time delay for specific points on the rings. 

To reduce the number of free parameters, we made the following simplifications: we assume that all three rings are intrinsically circular, and they are tilted only towards the line of sight (i.e., their normals are in x-z plane in Fig. 5). We also assume that every point on the rings moves ballistically from the time they were expelled from the centre, and we use a distance to the supernova of 50 kpc. We then fit tilted circular rings to the observed rings outlined in Fig. 1 by Crotts \& Heathcote (2000). In particular, we used the points a, b, c, and d in their figure and in our Fig. 5 (top), to constrain the geometry, because Crotts \& Heathcote (2000) provide detailed velocities from echelle spectra for these points. The velocities were $-$21.5, +22.3, +24.3, and +21.1 $\kms$ respectively, with regard to the supernova. These velocities correspond to positions along the line of sight (i.e., the x-axis in the lower panel of Fig. 5). We also used the coordinates where the slit in Maran et al. (2000) crossed the ORs to constrain the tilt of the rings. These are named M1 and M2 in Fig. 5. 
 
From our fits we find that the inclination angle of the ER is $\sim43^\circ$ and the inclination angles for the NOR and SOR are $\sim45^\circ$ and $\sim38^\circ$, respectively. These are consistent with previous meaurements. As can be seen in Fig. 5, both ORs appear to be shifted westwards relative to the ER. The general success of the circular fits can be seen at the positions a, b, c and d, as well as the positions where the slit used by Maran et al. (2000) crossed the rings, called M1 and M2 in Fig. 5. From these fits we estimate that the distances from the supernova to M1 and M2 are $\sim 1.65\times10^{18}$ cm, and $\sim 2.05\times10^{18}$ cm respectively.

The light-echo paraboloid is the surface in space with the supernova at its focus, on which all points have the same delay time since supernova breakout, $t_{\rm delay}$. The paraboloid swept over M1 (M2) roughly at $t_{\rm delay} \sim 260~(1183)$ days, respectively. In the analysis of Maran et al. (2000) it was assumed that the distance to both M1 and M2 was $2\times10^{18}$ cm, and that the difference in time delay was 950 days. 

\begin{table*}
\caption{Emission lines from the southern outer ring of SN1987A as observed with VLT-FORS1 in 2002 December}
\begin{tabular}{lccccc}
\hline
\hline
Emission line & Rest Wavelength (\AA) & Flux                                            & Velocity (km~s$^{-1}$)& Flux (j/jH$\beta$)& Extinction correction\\
\hline
${\rm [O~II]}$                 & 3726.0 + 3728.83          & 17.40(1.15)$^{\mathrm{a}}$  & N/A            & 1.115(0.110)   & 1.98$^{\mathrm{b}}$\\

H$\eta$          &3839.895                          &1.25(0.15)                                   &  300.10     &0.079(0.024)    &1.95\\
${\rm [Ne~III]}$             &3872.541                          &2.25(0.30)                                   &  290.00      &0.142(0.022)   &1.94\\
H$\zeta$        &3889.05    &2.30(0.40)  &  285.42       &0.145(0.027) &1.92\\
${\rm [Ne~III]}$      &3967.50    &0.85(0.15)  &  305.00       &0.053(0.010) &1.92\\
H$\varepsilon$  &3970.10    &2.75(0.35)  &  306.70       &0.174(0.040) &1.92\\
${\rm [S~II]}$        &4068.60    &4.90(0.40)  &   N/A	 &0.310(0.036) &1.90\\
${\rm [S~II]}$        &4076.35    &0.50(0.15)  &   N/A	 &0.032(0.010) &1.90\\
H$\delta$       &4101.73    &5.00(0.40)  &  317.65 	 &0.316(0.036) &1.89\\
H$\gamma$       &4340.50    &7.30(0.70)  &  305.49 	 &0.462(0.058) &1.84\\
${\rm [O~III]}$       &4363.21    &0.70(0.10)  &  306.50 	 &0.044(0.007) &1.84\\
He~II       &4685.70    &3.55(0.15)  &  290.03	 &0.224(0.020) &1.71\\
H$\beta$        &4861.32    &15.80(1.35) &  306.20 	 &1.000     &1.71\\
${\rm [O~III]}$       &4958.91    &3.90(0.45)  &  320.00 	 &0.246(0.035) &1.68\\
${\rm [O~III]}$       &5006.84    &15.60(1.30) &  298.90       &0.987(0.116) &1.67\\
${\rm [N~I]}$         &5200.30    &2.25(0.15)  &  260.76	 &0.145(0.050) &1.67\\
${\rm [N~II]}$        &5754.59    &2.35(0.30)  &  311.02       &0.148(0.022) &1.54\\
He~I       &5875.63    &6.30(0.50)  &  319.16       &0.398(0.045) &1.53\\
${\rm [O~I]}$         &6300.30    &2.90(0.40)  &  304.70       &0.183(0.029) &1.47\\
${\rm [N~II]}$        &6548.05    &26.70(1.10) &  312.82       &1.689(0.157) &1.45\\
H$\alpha$       &6562.80    &56.50(1.50) &  309.00       &3.580(0.655) &1.45\\
${\rm [N~II]}$        &6583.45    &79.75(1.80) &  301.46       &5.050(0.868) &1.45\\
${\rm [S~II]}$        &6716.44    &6.30(0.60)  &  306.77       &0.398(0.050) &1.45\\
${\rm [S~II]}$        &6730.82    &8.90(0.60)  &  303.35 	 &0.570(0.061) &1.43\\     
\hline
\end{tabular}
\begin{list}{}{}
\item[$^{\mathrm{a}}$] Flux is given in units of 10$^{-17}\ergsm$. 1-$\sigma$ uncertainties are given in brackets.
\item[$^{\mathrm{b}}$] Extinction correction.
\end{list}
\end{table*}
\begin{table*}
\caption{Balmer line ratios for the FORS1 and UVES data.}
\begin{tabular}{c c c c c c c c}
\hline
\hline
Line & FORS 1$^{\mathrm{a,b}}$ & UVES$^{\mathrm{c}}$ & UVES$^{\mathrm{d}}$ & UVES$^{\mathrm{e}}$ & UVES$^{\mathrm{f}}$ & UVES $^{\mathrm{g}}$ & Case B$^{\mathrm{h}}$\\
\hline
H$\alpha$      & 3.580  & 4.310 & 3.650  & 3.100 & 3.250 & 3.777 & 2.800\\
H$\beta$       & 1.000  & 1.000 & 1.000  & 1.000 & 1.000 & 1.000 & 1.000\\
H$\gamma$      & 0.462  & 0.153 & 0.219  & 0.221 & 0.196 & 0.402 & 0.466\\
H$\delta$      & 0.316  & 0.100 & 0.073  & N/A   & 0.088 & 0.159 & 0.256\\ 
H$\varepsilon$ & 0.174  & N/A   & N/A    & N/a   & N/A   & N/A   & 0.158\\
H$\zeta$       & 0.145  & N/A   & N/A    & N/A   & N/A   & N/A   & 0.105\\
H$\eta$        & 0.079  & N/A   & N/A    & N/A   & N/A   & N/A   & 0.073\\
\hline
\end{tabular}
\begin{list}{}{}
\item[$^{\mathrm{a}}$] All data are corrected for extinction using E$_(B-V)$ = 0.16 mag and R$_v$ = 3.1.
\item[$^{\mathrm{b}}$] Southern outer ring.
\item[$^{\mathrm{c}}$] Northern outer ring (2002).
\item[$^{\mathrm{d}}$] Southern outer ring (2002).
\item[$^{\mathrm{e}}$] Northern outer ring (2009).
\item[$^{\mathrm{f}}$] Southern outer ring (2009).
\item[$^{\mathrm{g}}$] South part of the inner ring (2002).
\item[$^{\mathrm{h}}$] For T = 10$^4$ K and N$_e$ = 10$^4$ cm$^{-3}$.
\end{list}
\end{table*}

\begin{table*}
\caption{Emission lines from the outer rings as observed with VLT-UVES in 2002 October}
\begin{tabular}{lcccccccc}
\hline
\hline
&&&North&&&South&&\\
& Rest wavel.& &$V_{peak}^{\mathrm{c}}$ & $V_{FWHM}$ & & $V_{peak}^{\mathrm{c}}$ & $V_{FWHM}$ & Extinction\\
Emission & (\AA) & Relative flux $^{\mathrm{a}}$ & ($\kms$) & ($\kms$) & Relative flux$^{\mathrm{a}}$ & ($\kms$) &($\kms$) &correction$^{\mathrm{b}}$\\
\hline
${\rm [Ne~V]}$&3425.86&$11.0\pm6.8$&$295.7\pm3.5$&$14.71\pm3.46$&$5.9\pm3.7$&$285.3\pm6.3$&$25.82\pm5.87$&2.06\\
${\rm [O~II]}$&3726.03&$-$&$-$&$-$&$19.2\pm2.8$&$292.2\pm0.7$&$11.21\pm0.51$&1.98\\
${\rm [O~II]}$&3728.82&$-$&$-$&$-$&$21.0\pm2.9$&$289.3\pm0.8$&$12.18\pm0.51$&1.98\\
${\rm [Ne~III]}$&3868.75&$22.2\pm5.3$&$282.1\pm1.9$&$19.86\pm1.69$&$18.9\pm10.7$&$288.7\pm5.2$&$22.33\pm5.06$&1.94\\
${\rm [S~II]}$&4068.60&$9.9\pm3.6$&$281.5\pm2.1$&$15.77\pm2.14$&$10.9\pm3.7$&$289.0\pm2.4$&$18.48\pm2.36$&1.90\\
${\rm [S~II]}$&4076.35&$-$&$-$&$-$&$-$&$-$&$-$&1.90\\
H$\delta$&4101.73&$10.0\pm4.0$&$282.8\pm4.4$&$28.97\pm4.35$&$7.3\pm3.2$&$287.8\pm2.6$&$16.18\pm2.75$&1.89\\
H$\gamma$&4340.46&$15.3\pm2.9$&$279.8\pm1.9$&$24.51\pm1.62$&$21.9\pm3.4$&$290.9\pm1.6$&$24.22\pm1.22$&1.84\\
${\rm [O~III]}$&4363.21&$10.3\pm3.2$&$276.6\pm4.3$&$36.61\pm4.78$&$5.3\pm1.5$&$290.8\pm2.0$&$18.70\pm1.94$&1.84\\
He~II &4685.7&$6.3\pm2.2$&$289.1\pm3.1$&$23.63\pm3.09$&$7.7\pm2.8$&$288.0\pm4.4$&$33.24\pm4.91$&1.75\\
H$\beta$&4861.32&$129.8\pm25.7$&$292.6\pm1.3$&$22.18\pm1.34$&$100.0$&$290.4\pm0.9$&$18.97\pm0.43$&1.71\\
${\rm [O~III]}$&4958.91&$69.6\pm14.0$&$288.6\pm2.4$&$32.12\pm3.07$&$48.6\pm5.4$&$291.0\pm1.0$&$19.90\pm0.54$&1.68\\
${\rm [O~III]}$&5006.84&$188.4\pm29.2$&$290.6\pm1.7$&$26.71\pm1.69$&$158.3\pm17.5$&$291.1\pm0.9$&$19.63\pm0.53$&1.67\\
${\rm [N~II]}$&5754.59&$13.8\pm2.2$&$287.5\pm1.8$&$25.67\pm1.31$&$10.8\pm1.3$&$290.2\pm0.8$&$14.21\pm0.49$&1.54\\
He~I&5875.63&$16.4\pm2.6$&$292.1\pm1.4$&$19.76\pm1.02$&$16.7\pm2.0$&$290.9\pm0.7$&$13.67\pm0.45$&1.53\\
${\rm [O~I]}$&6300.30&$14.3\pm1.6$&$285.6\pm1.0$&$19.62\pm0.58$&$11.8\pm1.3$&$291.1\pm0.5$&$10.44\pm0.25$&1.47\\
${\rm [O~I]}$&6363.78&$3.8\pm1.1$&$288.3\pm1.9$&$17.29\pm1.90$&$3.9\pm0.9$&$290.4\pm1.1$&$12.42\pm1.02$&1.47\\
${\rm [N~II]}$&6548.05&$179.4\pm17.4$&$294.0\pm0.7$&$20.18\pm0.31$&$198.7\pm17.7$&$291.9\pm0.2$&$10.28\pm0.06$&1.45\\
H$\alpha$&6562.80&$431.2\pm39.7$&$292.4\pm0.7$&$26.60\pm0.33$&$365.4\pm32.2$&$290.5\pm0.5$&22.25$\pm0.13$&1.45\\
${\rm [N~II]}$&6583.45&$572.9\pm53.3$&$292.9\pm0.6$&$20.08\pm0.24$&$620.3\pm54.9$&$290.6\pm0.2$&$10.32\pm0.05$&1.45\\
${\rm [S~II]}$&6716.44&$34.9\pm4.3$&$295.0\pm0.6$&$24.26\pm0.55$&$39.4\pm21.2$&$291.9\pm0.5$&$9.79\pm0.23$&1.43\\
${\rm [S~II]}$&6730.82&$52.8\pm8.4$&$294.8\pm0.7$&$17.01\pm1.14$&$58.0\pm5.4$&$291.6\pm0.3$&$9.66\pm0.12$&1.43\\
${\rm [Ar~III]}$&7135.79&$5.2\pm3.1$&$287.2\pm7.7$&$32.57\pm6.25$&$3.4\pm0.6$&$290.5\pm1.1$&$13.78\pm0.92$&1.39\\
\hline
\end{tabular}
\begin{list}{}{}
\item[$^{\mathrm{a}}$] Fluxes are given with regard to H$\beta$ for the SOR: ``100" corresponds to $(2.2\pm 0.2) \times 10^{-16}\ergsm$.
\item[$^{\mathrm{b}}$] $E(B-V)=0.16$, with $E(B-V)=0.10$ mag for LMC and $E(B-V)=0.06$ mag for the Milky Way.
\item[$^{\mathrm{c}}$] The recession velocity of the peak flux.
\end{list}
\end{table*}

\begin{table*}
\caption{Emission lines from the outer rings as observed with VLT-UVES in 2009 January}
\begin{tabular}{lcccccccc}
\hline
\hline
&&&North&&&South&&\\
 & Rest wavel.& &$V_{peak}^{\mathrm{c}}$&$V_{FWHM}$&&$V_{peak}^{\mathrm{c}}$&$V_{FWHM}$&Extinction\\
Emission&($\AA$)&Relative flux$^{\mathrm{a}}$&($\kms$)&($\kms$)&Relative flux$^{\mathrm{a}}$&($\kms$)&($\kms$)&correction$^{\mathrm{b}}$\\
\hline
${\rm [Ne~V]}$&3425.86&$45.6\pm27.1$&$284.3\pm11.5$&$46.76\pm11.04$&$30.4\pm12.4$&$285.5\pm3.7$&$23.67\pm3.92$&2.06\\
${\rm [O~II]}$&3726.03&$-$&$-$&$-$&$45.2\pm9.4$&$290.6\pm1.1$&$12.08\pm0.93$&1.98\\
${\rm [O~II]}$&3728.82&$-$&$-$&$-$&$38.4\pm6.1$&$288.1\pm1.0$&$13.24\pm0.75$&1.98\\
${\rm [Ne~III]}$&3868.75&$37.2\pm9.2$&$283.1\pm1.6$&$16.34\pm1.52$&$25.9\pm7.2$&$288.2\pm2.4$&$22.20\pm2.35$&1.94\\
${\rm [S~II]}$&4068.60&$-$&$-$&$-$&$-$&$-$&$-$&1.90\\
${\rm [S~II]}$&4076.35&$-$&$-$&$-$&$-$&$-$&$-$&1.90\\
H$\delta$&4101.73&$-$&$-$&$-$&$8.8\pm5.2$&$294.9\pm6.9$&$28.98\pm6.66$&1.89\\
H$\gamma$&4340.46&$25.5\pm9.9$&$279.0\pm8.1$&$54.7\pm9.1$&$19.6\pm3.7$&$291.6\pm1.9$&$23.66\pm1.65$&1.84\\
${\rm [O~III]}$&4363.21&$-$&$-$&$-$&$-$&$-$&$-$&1.84\\
He~II &4685.7&$9.5\pm3.3$&$295.3\pm3.0$&$22.53\pm3.02$&$5.3\pm3.0$&$285.6\pm6.1$&$28.19\pm5.89$&1.75\\
H$\beta$&4861.32&$115.1\pm23.2$&$292.3\pm2.2$&$36.04\pm4.88$&$100.0$&$290.6\pm1.0$&$21.46\pm0.54$&1.71\\
${\rm [O~III]}$&4958.91&$69.2\pm10.3$&$290.5\pm1.7$&$26.21\pm1.65$&$73.6\pm6.7$&$291.2\pm0.9$&$19.66\pm0.50$&1.68\\
${\rm [O~III]}$&5006.84&$237.1\pm26.1$&$287.9\pm1.4$&$28.53\pm1.34$&$227.7\pm20.0$&$290.9\pm0.9$&$19.64\pm0.47$&1.67\\
${\rm [N~II]}$&5754.59&$10.6\pm2.8$&$288.9\pm2.1$&$20.70\pm2.03$&$11.3\pm2.0$&$290.9\pm0.9$&$11.43\pm0.75$&1.54\\
He~I&5875.63&$9.6\pm1.5$&$294.7\pm1.1$&$15.15\pm0.86$&$13.1\pm1.8$&$290.8\pm0.9$&$13.72\pm0.65$&1.53\\
${\rm [O~I]}$&6300.30&$12.5\pm1.8$&$288.4\pm1.5$&$21.63\pm1.11$&$10.6\pm1.1$&$291.6\pm0.5$&$8.90\pm0.27$&1.47\\
${\rm [O~I]}$&6363.78&$4.4\pm1.9$&$286.3\pm3.5$&$21.21\pm3.38$&$4.1\pm1.1$&$291.8\pm1.2$&$11.44\pm1.09$&1.47\\
${\rm [N~II]}$&6548.05&$134.3\pm10.4$&$294.5\pm0.6$&$19.99\pm0.32$&$155.2\pm10.8$&$291.9\pm0.3$&$10.88\pm0.09$&1.45\\
H$\alpha$&6562.80&$356.1\pm29.4$&$292.1\pm0.9$&$28.57\pm0.67$&$325.5\pm24.9$&$290.3\pm0.6$&$22.60\pm0.16$&1.45\\
${\rm [N~II]}$&6583.45&$439.1\pm31.9$&$292.9\pm0.5$&$19.36\pm0.26$&$477.6\pm32.6$&$290.6\pm0.3$&$10.84\pm0.08$&1.45\\
${\rm [S~II]}$&6716.44&$24.9\pm4.2$&$296.3\pm0.7$&$10.06\pm0.83$&$30.1\pm2.4$&$291.5\pm0.4$&$10.57\pm0.20$&1.43\\
${\rm [S~II]}$&6730.82&$34.8\pm6.0$&$295.7\pm0.8$&$13.41\pm1.90$&$39.4\pm3.2$&$291.8\pm0.4$&$9.36\pm0.18$&1.43\\
${\rm [Ar~III]}$&7135.79&$5.6\pm1.4$&$289.8\pm2.5$&$25.00\pm2.39$&$4.3\pm0.7$&$289.9\pm0.9$&$11.09\pm0.69$&1.39\\
\hline
\end{tabular}
\begin{list}{}{}
\item[$^{\mathrm{a}}$] Fluxes are given in respect to H$\beta$ for the SOR: ``100" corresponds to $(2.5\pm 0.2) \times 10^{-16}\ergsm$.
\item[$^{\mathrm{b}}$] $E(B-V)=0.16$, with $E(B-V)=0.10$ mag for LMC and $E(B-V)=0.06$ mag for the Milky Way.
\item[$^{\mathrm{c}}$] The recession velocity of the peak flux.
\end{list}
\end{table*}

For our points T1 (area 4) and T2 (area 1), we find distances to the supernova of $\sim 1.92\times10^{18}$  cm and $\sim1.85\times10^{18}$ cm, respectively, and the delay times, as marked by the light-echo paraboloids in Fig. 5, are $\sim 620$ and $\sim 768$ days, respectively. The Doppler shifts with regard to the supernova at these points in the model are $-2$ and $+4$ $\kms$, respectively. For areas 2 and 3 in Fig. 1 and Table 9, the distances to the supernova are $\sim 2.06\times10^{18}$ cm and $\sim 1.80\times10^{18}$ cm, respectively, and the delay times are $\sim 648$ and $\sim 604$ days, respectively. The Doppler shifts with regard to the supernova are $-4$ and $-2$ $\kms$, respectively. In the geometrical model, the distance between the ORs and the supernova range from $(1.73-2.15)\times10^{18}$ cm for the NOR and $(1.79-2.03)\times10^{18}$ cm for the SOR. The shortest distance occurs for the south-eastern part of the NOR, just north of the projected intersection with the SOR.

\subsection{Analysis of the spectroscopy}

Because of the low S/N and the proximity between the ER and the NOR location "area 1", no FORS1 spectrum was extracted from this position. The results from the SOR are shown in Table 4. For the SOR we were able to measure the Balmer lines and extract important information about the plasma diagnostics from various other lines (i.e. [N~II], [O~III], and [S~II]). The results from the Balmer lines are shown in Table 5 with the theoretical values of case B at 10$^4$ K, for an electron density of 10$^4$ cm$^{-3}$ (Osterbrock \& Ferland 2006). In general the measured ratios are close to the case B values, except for the higher j$_{H\alpha}$/j$_{H\beta}$ ratio. This difference can be due to collisional excitation of H$\alpha$ (Lundqvist \& Fransson 1996). On the other hand, our estimated ratio of j$_{H\epsilon}$/j$_{H\beta}$, which is also higher than the prediction of case B, is caused by mixing with [Ne~III] $\lambda$3968, while the j$_{H\zeta}$/j$_{H\beta}$ ratio is affected by He I $\lambda$3889. In Table 5 we show all measured Balmer lines ratios corrected for extinction for both the FORS 1 and UVES spectra with measurements from the southern region of the ER from the 2002 epoch for comparison.  

The ratios of forbidden emission lines can be used to estimate temperature and electron density in the ORs of SN 1987A. For both spectroscopy data sets we measured the emission lines and made estimates of the temperature and electron density. For the FORS1 data, we measured the line fluxes in our spectra with IRAF's task SPLOT. The most important source of uncertainty is the spectral resolution. Based on the spatial profile from [N~II]~$\lambda$6583 (cf. Fig. 3), it is evident how difficult it is to distinguish the profile of the SOR from the profile of the south part of the ER. The low resolution of the FORS1 spectra was also evident in the flux measurements of the Balmer lines. In Table 4 we show the fluxes we measured for the various emission lines along with the 1$\sigma$ errors that were calculated with the help of the SPLOT tool in IRAF. The applied extinction correction for each line is also shown in Table 4. The error was measured by using more pixels for the line profile and considering the additional noise as the uncertainty measurement.

The gas temperatures and densities in the ORs were estimated in the same way as in Gr\"oningsson et al. (2008), treating N$^+$ and O$^{2+}$ as six-level atoms, and O$^+$ and S$^+$ as five-level atoms. From Fig. 6 for the 2002 FORS1 data, we see that the [O~III] temperature on day 5791 for the SOR ("area 4") was $\sim 2.5\times10^4$~K, and that the [N~II] temperature was in the range $(1.3-1.4)\times10^4$~K. For the electron density, $N_{\rm e}$, we find $N_{\rm e} \lsim 3.3\times10^3$~cm$^{-3}$ from [S II]~$\lambda\lambda$6716,6731, assuming that the temperature of the emitting gas is not higher than that of the emitting [N~II]. The S$^+$ ion provides a second ratio using j$_{\lambda\lambda4069,4076}$/j$_{\lambda\lambda6716,6731}$. The blue lines come from hotter gas than [S~II]$\lambda\lambda$6716, 6731, but it is still somewhat surprising that the overlap between the j$_{\lambda\lambda4069,4076}$/j$_{\lambda\lambda6716,6731}$, and the ratio of the two red lines alone occurs at a temperature more characteristic of the [O~III] lines. What seems most plausible is that we have underestimated the errors of the relatively weak blue [S~II] lines. 
 
In a similar manner from the roughly coeval (i.e., days $5704-5705$) UVES data (listed in Table 6) of the SOR, we got an [O~III] temperature in the range $(1.95-2.00)\times10^4$~K, and a [N~II] temperature in the range $(1.00-1.10)\times10^4$~K. For the electron density, $N_{\rm e}$, we find $N_{\rm e} \lsim 1.0\times10^3$~cm$^{-3}$ from [O~II] using the [O~III] temperature as an upper limit to the [O~II] temperature. The [S~II] density could not be constrained. For 2009 (days $7944-8021$, see Table 7) the numbers are $(1.15-1.25)\times10^4$~K, $N_{\rm e} \lsim 3.0\times10^3$~cm and $N_{\rm e} \lsim 2.3\times10^3$~cm$^{-3}$ for [N~II], [S~II], and [O~II], respectively, assuming the [O~III] temperature is still $\sim 2\times10^4$ K. The agreement for [N~II] and [S~II] between our estimates using FORS1 and UVES for the 2002 epoch makes us believe that a [N II] temperature slightly in excess of $10^4$~K is correct, as well as a [S~II] density below $\sim 3\times10^3$~cm$^{-3}$. The [O~III] temperature is most likely in excess of $2\times10^4$~K, and close to $2.5\times10^4$ K, as the relative fluxing of the UVES data in the blue region of, e.g., [O~III]~$\lambda$4363, gives worse fits to Balmer line ratios for the blue hydrogen lines than do the FORS 1 data. The higher derived [O~II] density for the 2009 epoch reflects a larger uncertainty in the [O~II] line fluxes than for the 2002 epoch. We find no reason to believe the [O~II] density in reality should be much higher in 2009 than in 2002. 

The UVES data also make it possible for us to make estimates for the NOR ring ("area 1"). For the 2002 data, we got a [O~III] temperature of $\sim 2.7\times10^4$~K, and a [N~II] temperature of $\sim 1.2\times10^4$~K. For the electron density, $N_{\rm e}$, the result for [S~II] is consistent with the 2002 result, but a good upper limit is impossible to obtain. No density estimate from [O~II] was possible. In 2009 the derived value for [N~II] is still $\sim 1.2\times10^4$~K, whereas the error is too large to constrain the density. 
 
All results from our plasma diagnostics and the previous estimates by others can be seen in Table 8. Maran et al. (2000) reported an electron density $\lsim 2.2(1.6)\times10^{3}$~cm$^{-3}$ for the NOR (SOR) on day 4282 using [S~II]. Panagia et al. (1996) reported $\sim 800$ cm$^{-3}$ for day 2877 using [O II] for the northern part of the NOR (with a light-travel delay time of $\sim 660$ days according to the model in Fig. 5) and the southern part of the SOR (with an estimated light-travel delay time of $\sim 685$ days). Panagia et al. (1996) estimated an error of $\sim 25$\% for the density, which would give an upper limit of $\sim 10^3$ ~cm$^{-3}$. As indicated by Fig. 7, this limit could be slightly higher if the temperature characterizing the [O~II] emission is higher than the $1.25\times10^4$ K assumed by Panagia et al. To be consistent with our estimates, we put a firm upper limit for the [O~II] temperature equal to that of [O~III], and arrive at $\lsim 1.3\times10^3$~cm$^{-3}$. In preliminary time-dependent photoionisation model results presented in Lundqvist (2007), we estimated a density for the [O~III]~$\lambda\lambda$4959, 5007 emitting gas for the 2002 epoch discussed in this paper to be just in excess of $10^3$ ~cm$^{-3}$ and for [O~III]~$\lambda$4363 just below $10^3$~cm$^{-3}$. The red [S~II] lines required a density around $\sim 3\times10^3$~cm$^{-3}$. These numbers generally agree with the nebular analysis presented here, as well as with Panagia et al. (1996), and Maran et al. (2000). Note that in Table 8 we included estimates of the light-travel time delay using the geometric model discussed in Sect. 4.1. Although the full analysis of model calculations using various prescriptions for the flash is postponed for future work, we note that Table 8 shows that there is no obvious decrease in the density of the emitting gas between days $\sim 2200$ and $\sim 7200$ (as corrected for light travel time).

\begin{table*}
\caption{Compilation of plasma diagnostics for the outer rings of SN1987A}
\begin{tabular}{l c c c c c c}
\hline
\hline
Epoch & Corrected epoch$^{\mathrm{a}}$ & Ring & $N_{e}$ (cm$^{-3}$) & $T$ (K) & Diagnostic & Reference\\
\hline
2877& 2192 & SOR & $<$ 1.3$\times$10$^{3}$$^{\mathrm{b}}$ &  & [O~II] & Panagia et al. (1996)\\
2877& 2217 & NOR & $<$ 1.3$\times$10$^{3}$ &  & [O~II] & Panagia et al. (1996)\\
4282& 3099 & NOR & $<$ 2.2$\times$10$^{3}$ &  & [S~II] & Maran et al. (2000)\\
4282& 4022 & SOR & $<$ 1.6$\times$10$^{3}$ &  & [S~II] & Maran et al. (2000)\\
5702--5705& 4902--4905 & SOR & $<$ 1.0$\times10^{3}$ &  & [O~II] & This paper\\
5702--5705& 4902--4905 & SOR &  & $(1.00-1.10)\times10^4$ & [N~II] & This paper\\
5702--5705& 4902--4905 & SOR & & $(1.95-2.00)\times10^4$ & [O~III] & This paper\\
5702--5705& 4934--4937 & NOR &  & $\sim1.2\times10^4$ & [N~II] & This paper\\
5702--5705& 4934--4937 & NOR & & $\sim2.7\times10^4$ & [O~III] & This paper\\
5791& 4991 & SOR & $<$ 3.3$\times10^{3}$ &  & [S~II] & This paper\\
5791& 4991 & SOR &  & $(1.3-1.4)\times10^4$ & [N~II] & This paper\\
5791& 4991 & SOR & & $\sim2.5\times10^4$ & [O~III] & This paper\\
7944--8021& 7144--7221 & SOR & $<$ 2.3$\times10^{3}$ &  & [O~II] & This paper\\
7944--8021& 7144--7221 & SOR & $<$ 3.0$\times10^{3}$ &  & [S~II] & This paper\\
7944--8021& 7144--7221 & SOR &  & (1.15-1.25)$\times10^4$ & [N~II] & This paper\\
7944--8021& 7176--7253 & NOR &  & $\sim1.2\times10^4$ & [N~II] & This paper\\
\hline
\end{tabular}
\begin{list}{}{}
\item[$^{\mathrm{a}}$] In days. Corrected for light-travel time.
\item[$^{\mathrm{b}}$] Allowing for a higher temperature than Panagia et al. (1996).
\end{list}
\end{table*}

\begin{figure*}[ht*]
\centering
\includegraphics[width=9cm]{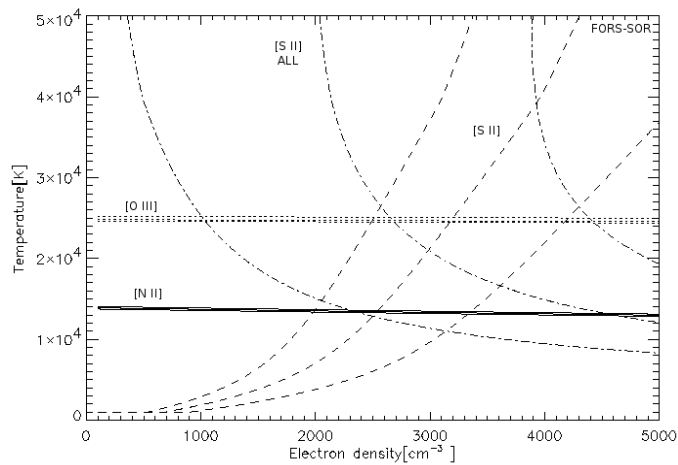}
\caption{Electron density vs. temperature for dereddened line ratios from the SOR. The data are from the 2002 FORS1 spectrum. The shown lines are: [N~II] $\lambda$$\lambda$6548,6581/$\lambda$5755 (solid), [O~III] $\lambda$$\lambda$4959,5007/$\lambda$4363 (dotted), [S~II] $\lambda$$\lambda$4069,4075/$\lambda$6716,6731 (dotted-dashed), and [S II] $\lambda$6716/$\lambda$6731 (dashed). Each ratio is represent by three curves where the outer curves are for the 1$\sigma$ statistical uncertainty.}
\end{figure*}

\begin{figure*}[ht]
\centering
\includegraphics[width=9cm]{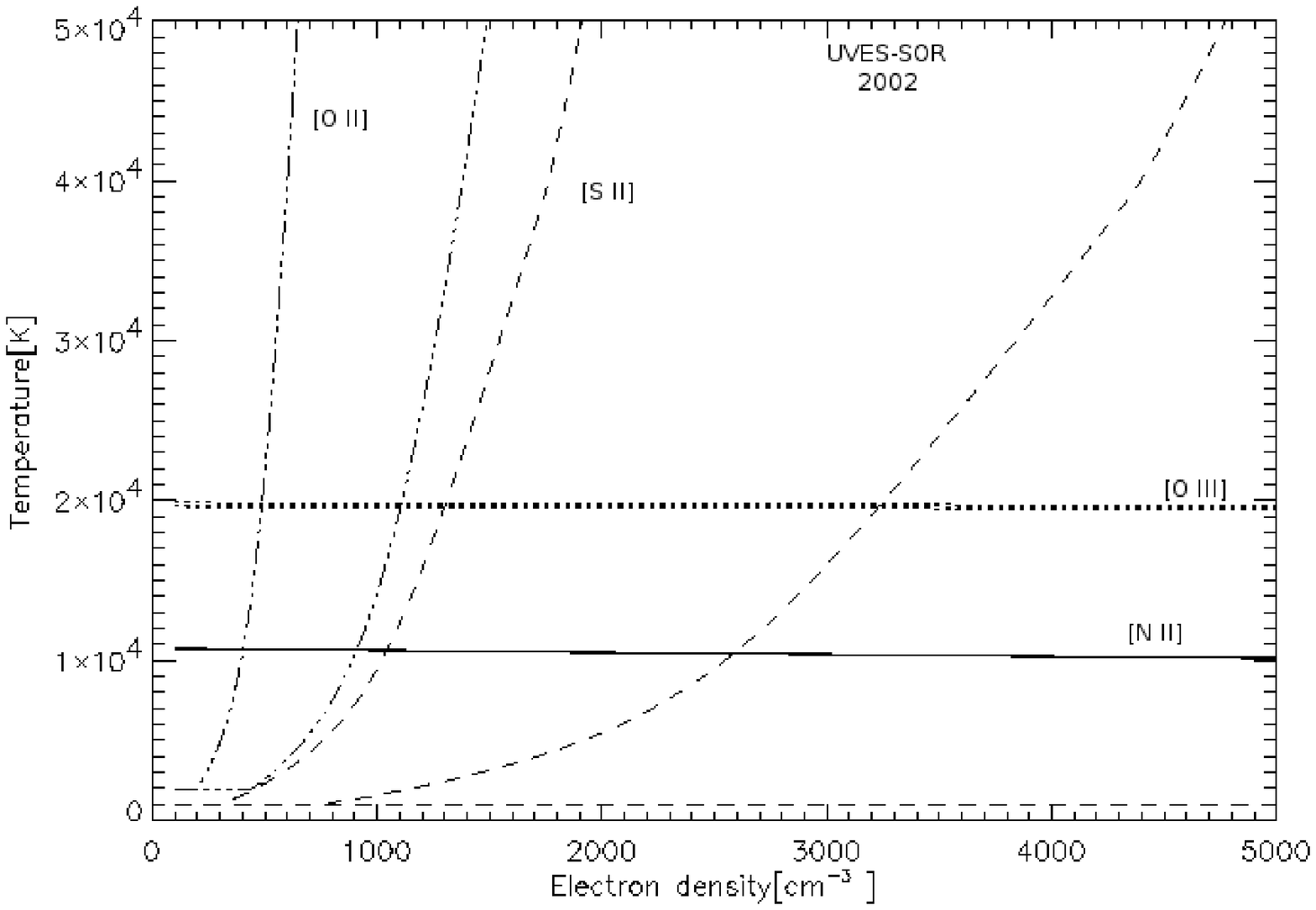}\hspace*{1mm}\includegraphics[width=9cm]{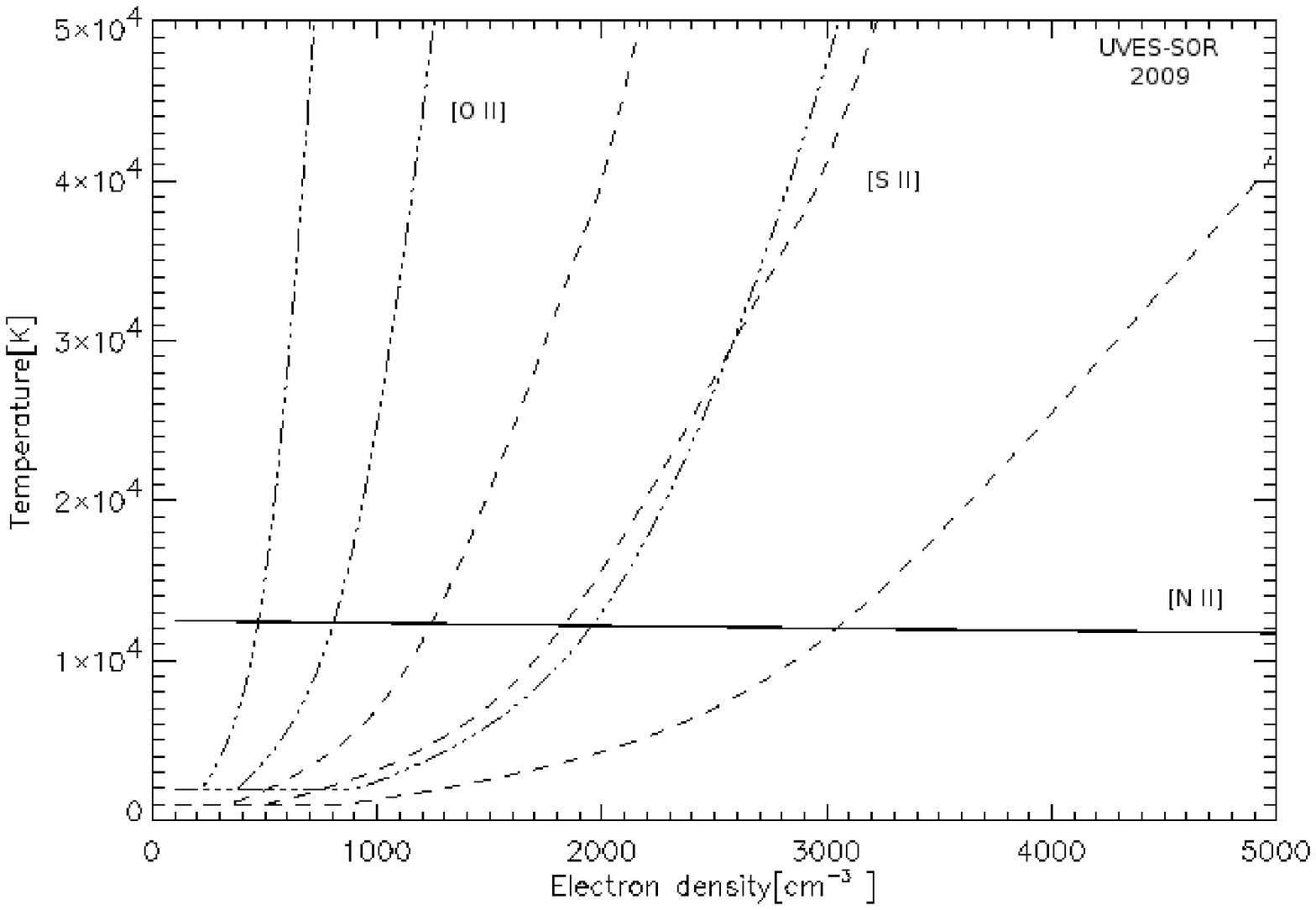}
\caption{Electron density vs. temperature for dereddened line ratios of the SOR of SN 1987A, using UVES data. The left panel shows the data from the 2002 epoch, while the right profile the 2009 epoch. In both panels the shown lines are: [N~II] $\lambda$$\lambda$6548,6581/$\lambda$5755 (solid), [O~III] $\lambda$$\lambda$4959,5007/$\lambda$4363 (dotted), [S II] $\lambda$6716/$\lambda$6731 (dashed) and [O II] $\lambda$$\lambda$3726,3729 (double-dotted dashed).}
\end{figure*}

\begin{figure*}[ht]
\centering
\includegraphics[width=9cm]{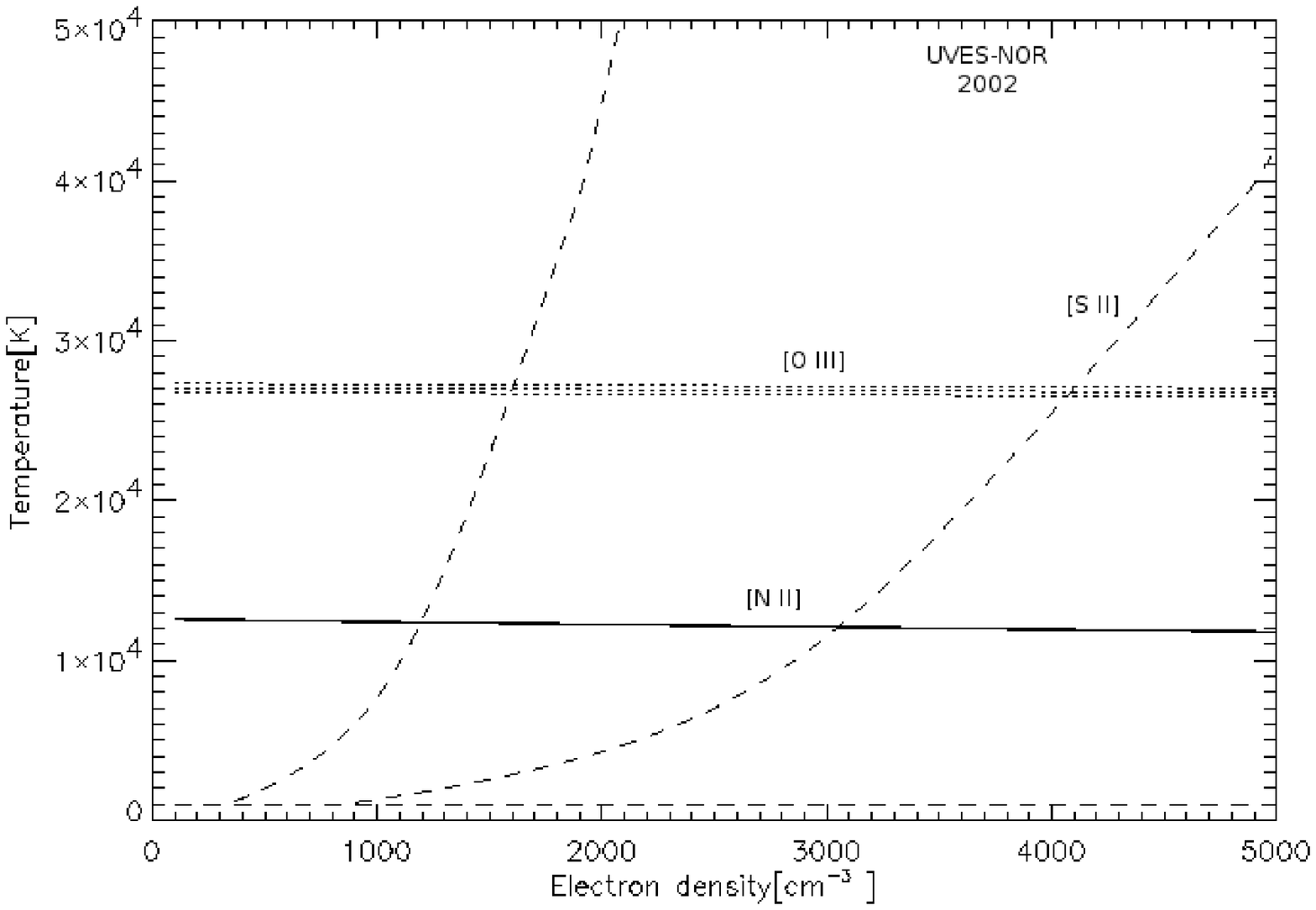}\hspace*{1mm}\includegraphics[width=9cm]{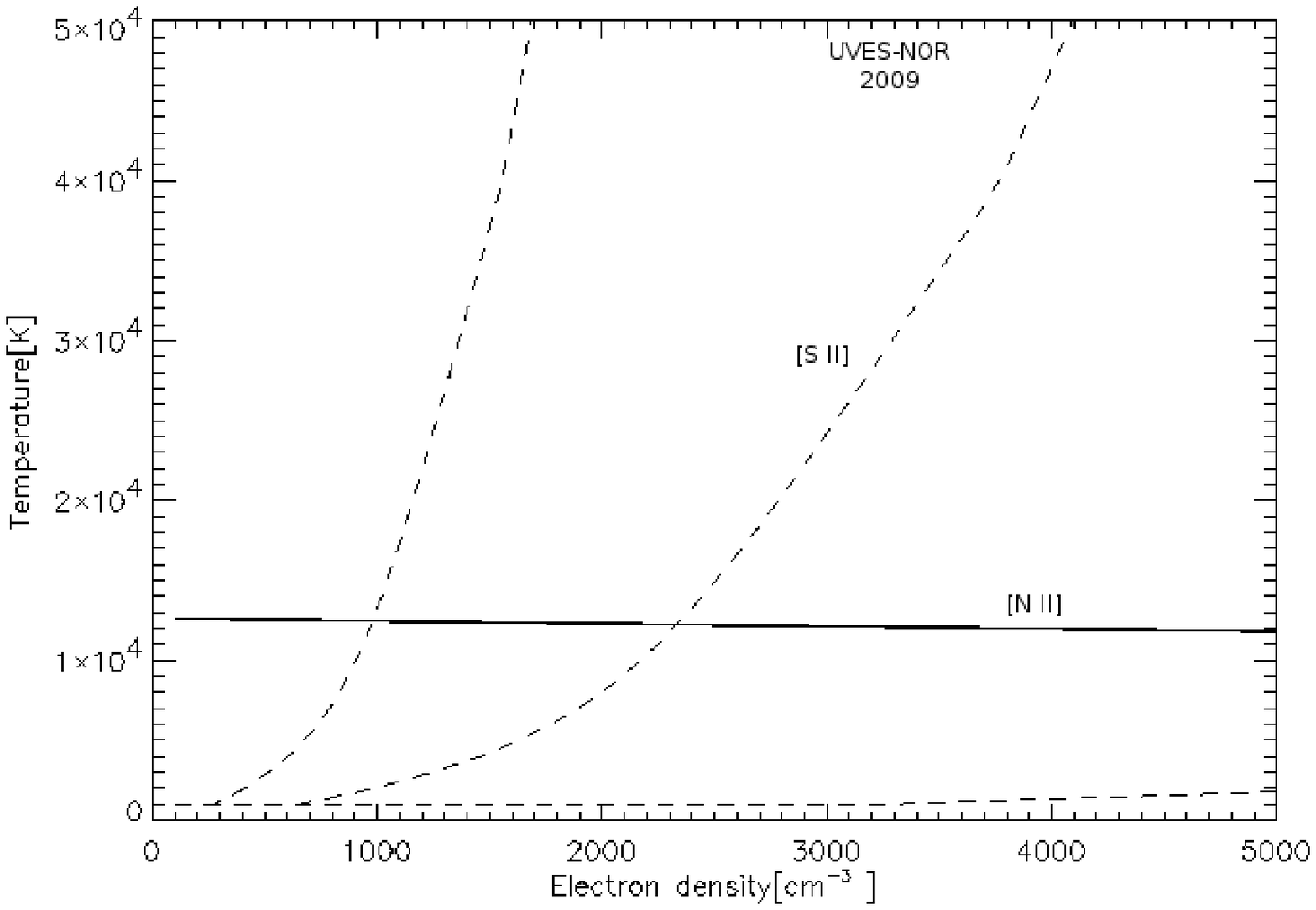}
\caption{Electron density vs. temperature for dereddened line ratios of the NOR of SN 1987A, using UVES data. The left panel shows the data from the 2002 epoch, while the right profile the 2009 epoch. In both panels the shown lines are: [N~II] $\lambda$$\lambda$6548,6581/$\lambda$5755 (solid), [O~III] $\lambda$$\lambda$4959,5007/$\lambda$4363 (dotted), and [S II] $\lambda$6716/$\lambda$6731 (dashed)}
\end{figure*}

Looking at the evolution of the absolute fluxes of the UVES data, we see that the emission in H$\alpha$, [N~II], and [S~II] decays with time (see Tables 6 and 7). As noted above, the densities of the various line-emitting gases (i.e., [O~II] and [S~II]) remain fairly constant, in a flash-ionisation scenario, wherefore one could expect the fluxes of those lines to decrease. For the [O~III] and [O~II] doublets, we note an increase though. This could be real, but a difficulty is that the increasing emission from the collision of the supernova ejecta and the ER could spill over into the ORs measurements using ground-based data. In Sect. 4.3 we test this using HST data. 

\subsection{Analysis of the photometry}

The plethora of HST data of SN 1987A in the archives provides an opportunity for investigating the evolution of the ORs in detail. Using various narrow band emission line filters from WFPC2 and ACS, we were able to construct light curves for specific emission lines (i.e., [O~III], [N~II], and H$\alpha$. This was done by performing surface brightness photomety using IDL routines. Our first consideration was to make sure that we measure the flux in the exact same areas in every image. For this, all our images were aligned with subpixel accuracy. Then the images were transformed to polar coordinates, and we measured the flux in the areas of interest using circular apertures. In total we chose four areas based on two criteria. First, we wanted areas that are not near the two bright neighbouring stars, and we also avoided the regions that are close to the ER. Areas near the two bright neighbouring stars are contaminated by light leaks from them, which increases the uncertainty in the photometric measurements. Areas 1 and 4 correspond to the same areas that were covered by the slit of the spectroscopic observations with UVES and FORS 1. As a cross-check for remaining background effects, we chose many regions near the supernova remnant on every image, and measured an average value of the remaining background which was subtracted from the measured flux. An interesting feature is the difference in the flux between regions just outside the ORs and the inter-ring region of SN 1987A. On average, a difference of 20$\%$ was measured in the flux of these regions. This effect can be caused by material that sits between the ER and the ORs, but it can also be caused by reflections from the nebula itself. An effect like this was also noted and measured by L. Wang (private communication, see Lundqvist 1999) using HST data.

Apart from monitoring the ORs, the photometric data were also used as a cross-check for the spectroscopic measurements using UVES \& FORS1. Small differences in the fluxes between photometry and spectroscopy can be caused by a small uncertainty in the position from which the photometric measurements were made with respect to spectroscopy. 

To compare the results from photometry with those from spectroscopy, we took the transmission curves of the HST filters and plotted their profiles to see which lines they include. The curve was obtained from the STSDAS package SYNPHOT within IRAF and is shown in Fig. 9. For the F502N filter [O~III] from WFPC2, the situation is rather simple because there is no inteference with any nearby line. For the F502N ACS [O~III] filter, there is contribution in the blue part of the filter from the [O~III]~$\lambda$4959. For the H$\alpha$ and the [N II] filters, the situation is slightly more complicated. As can be seen in Fig.~9, [N~II] $\lambda$6548 and H$\alpha$ are both included in the F656N WFPC2 filter. The situation is similar with the F658N filter, because there is contamination from the blue part of the H$\alpha$ line. Even though the filter does not include the entire H$\alpha$ line (we estimated a $\sim$30\% contribution), it cannot be considered negligible, and we took this into account in our analysis. On the other hand, the F658N ACS filter includes both the H$\alpha$ and the [N~II] doublet. The plot in Fig. 9 was also used to estimate the evolution of the emission line ratios between the [N~II]~$\lambda\lambda$6548, 6583 doublet and H$\alpha$. This can be done because the ratio between the [N~II] doublet has a constant value of three. Another assumption we made is that we consider the changes in the line ratios within a few months to be negligible. This allowed us to use observations with different dates. However, we were unable to determine the line ratio of the [N~II] doublet over H$\alpha$ for all epochs, especially for the 1998 and 1999 epochs. 

Starting with the SOR and the 1994 epoch, we measured the line ratio of the [N~II] doublet over H$\alpha$ to be $\sim 2.00$. This increased slightly to $\sim 2.10$ in the 1996 epoch. The next epoch where we measured the line ratio was for the 2002 epoch. From the FORS1 data we found the line ratio to be $\sim 1.90$, while based on the UVES data we found a value of $\sim 2.25$. This difference arises because we were unable to distinguish the SOR as clearly from the ER in the FORS1 data as we were for the UVES data. For the FORS1 data we used a smaller aperture to extract our spectrum than for UVES. Similarly for the NOR, for the 1994 epoch we estimated a ratio of $\sim 2.05$. Again, we found that the ratio increased to $\sim 2.13$ for the 1996 epoch, and moved up to $\sim 2.27$ for the 2001 epoch and further to $\sim 2.37$ for the 2003 epoch. Owing to the lack of WFPC2 data for 2004 and 2005, no direct estimates of the [N II]/H$\alpha$ line ratio could be made. Based on the evolution of the line ratio of the [N~II]~$\lambda\lambda$6548, 6583 doublet over H$\alpha$, we believe that the value for the 1994 epoch is slightly over-estimated. The high uncertainties of the [O~III] measurements did not allow us to derive a trend for the [O~III]/H$\alpha$ ratio.

In general, we did not come across any major problems during our photometric analysis. In some cases though, some of the [O~III] images suffered from reflections caused by the two neighbouring stars. The affected areas were not included in our analysis. All photometric data were corrected for extinction with the same extinction law as for the spectra. In Table 9 we present our entire photometric results from the HST data with the uncertainties. For the [O~III] images, the uncertainty is $\sim$ 30\% on average, while for the H$\alpha$ and [N~II] filters, the uncertainty is on average $\sim$ 20-25 \% (depending on the epoch), and $\sim$ 20 \%, respectively. 

In Fig. 10 we show our light curves for H$\alpha$, [N~II], and [O~III]. The time on the x-axis is $t_{\rm corr} = t - t_{\rm delay}$, i.e., the light-travel time corrected epoch as discussed in Sect. 4.1 and 5, where $t_{\rm delay}$ is 768 days for area 1 and 800 days for area 4. All light curves are for areas 1 and 4, i.e., the two areas that the spectroscopic slit from both FORS1 and UVES covered. In addition, these two regions did not suffer from reflections from the neighbouring stars (at least in H$\alpha$ and [N II]). As seen in Fig.~10, the shapes of the H$\alpha$ line and [N~II] doublet light curves decay as expected. The [O~III] doublet light curves show inconsistencies, most likely because of contamination from the neighbouring stars. Starting with the light curve from the SOR (see Fig. 10, upper left panel), we see that the [O~III] flux from day 5816 shows an increase compared with the last WFPC2 epoch (on day 5410). Because the effect was not large, we decided to use this flux in the light curve. Finally, the measurement from the last ACS epoch of the [O~III] data (day 6512) suffered severely from reflections, and it was excluded from the plot. The [O~III] doublet light curve from the NOR (see Fig. 10, upper right panel) suffers from the same reflections problems. The last two ACS epochs (days 6130 and 6512) were highly affected by the neighbouring stars, and we decided just for showing the effect to include the point from day 6512 in our light curve. 

Just as for the inner ring (cf. Lundqvist \& Fransson 1996, Mattila et al. 2010), the decay of the H$\alpha$, [N~II] and [O~III] emission is a combination of recombination and cooling. We can gain insight into this by studying the decay times for this emission. To do this, we created fits to the line light curves with the function $f_{\rm line} = C exp(-t_{\rm corr}/t_{\rm efold})$, where $t_{\rm efold}$ is the decay time of the line flux. For H$\alpha$, the decay times in Fig. 10 are $\sim 2.8\times10^3$ days and $\sim 4.2\times10^3$ days for areas 4 and 1, respectively. For [N~II] they are $\sim 3.2\times10^3$ days and $\sim 4.6\times10^3$ days for areas 4 and 1, respectively, and for [O~III] they are $\sim 2.9\times10^3$ days and $\sim 6.8\times10^3$ days, respectively. In general, the fluxes fall faster in the southern outer ring (area 4) than in the northern (area 1). The long decay time for [O~III] in area 1 should be taken with caution, as the [O~III] ACS data from 2003 and 2004 are uncertain. More recent data are needed to see whether the upturn in [O~III] in area 1 after $t_{\rm corr} \gsim 5000$ days from HST photometry and from VLT/UVES spectroscopy (see Sect. 4.2) is real or not.

Although H$\alpha$ is expected to get some contribution from collisional excitation (cf. Sect. 4.2), H$\alpha$ is to a large fraction caused by recombination emission. For a temperature of $10^4$ K, the case B recombination time of hydrogen is $t_{\rm rec,H} \approx 1.5\times10^4 (N_e/3\times10^3 \cm3)^{-1}$ days, which is 3--5 times longer than the decay time of the H$\alpha$ emission from the ORs in Fig. 10. With an He/H-ratio of $\sim 0.2$ by number, as for the ER (Mattila et al. 2010), the number density of free electrons during recombination is also reduced by recombination of fully ionised helium, which for $10^4$ K has a recombination time of $t_{\rm rec,He} \approx 0.34 t_{\rm rec,H}$. Recombination of fully ionised helium can, however, only contribute $\sim 20\%$ of the loss of free electrons on a time scale comparable to the H$\alpha$ decay time. To account for the H$\alpha$ decay, we also need to invoke fast decay of the collisional excitation part of H$\alpha$ as well as a density exceeding $3\times10^3 \cm3$. The former rapidly occurs when the gas cools below $\sim 1.5\times10^4$ K. The upper limit on density can only be found through modelling similar to that for the ER (cf. Mattila et al. 2010), but it seems hard to avoid that gas with density components initially with $N_e \sim 5\times10^3 \cm3$ may be needed. There is no contradiction in that the electron density in the [S~II] emitting gas is lower than $N_e \sim 3\times10^3 \cm3$ (cf. Table 8) because the gas in this region only needs to be partially ionised.

The decay times of the [N~II] emission for the ORs are similar to those of H$\alpha$. For [N~II], this is not mainly because of recombination. In a $3\times10^3 \cm3$ model of Mattila et al. (2010) for the ER, nitrogen is dominated by N$^+$ between $\sim (3-9)\times10^3$ days with only minor changes in the respective fraction of N$^+$/N. However, during the same period, the temperature drops from $\sim 1.5\times10^4$ K to $\sim 4\times10^3$ K in the N$^+$ gas. Whereas H$\alpha$ recombination emission is fairly insensitive to temperature, the collisionally excited [N~II] is not. So while a density component with $3\times10^3 \cm3$ can radiate H$\alpha$ at late times, it is inefficient in emitting [N~II]. That we spectroscopically measure $\sim 1.1\times10^4$ K for [N~II] means that the density components that have temperatures in this range will dominate the emission of [N~II]. For the ER, the $3\times10^3 \cm3$ model obtains that temperature after $\sim 3.5\times10^3$ days. Because the evolution of different density components scales roughly inversely with density, components with density around $1.5\times10^3 \cm3$ should dominate [N~II] after $7\times10^3$ days, which is consistent with our upper limits on density from [S~II] and [O~II] in Table 8. So while H$\alpha$ can give a hint on the density components with the highest density, lines like [N~II] are useful for estimating the relative distribution of mass in different density components.

\begin{figure}[ht]
\centering
\includegraphics[width=9cm]{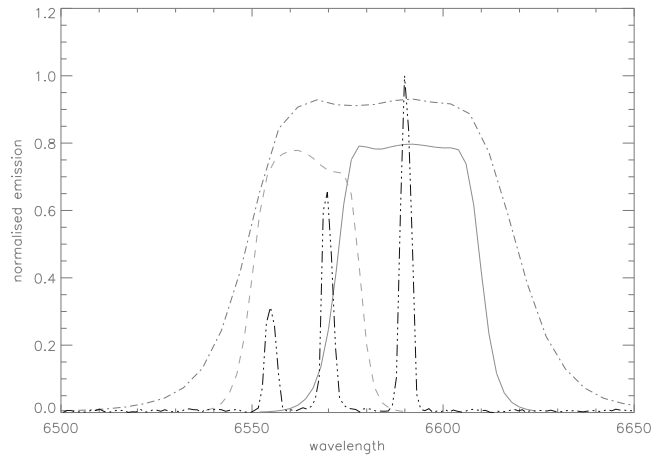}
\caption{Filter and line profiles for SN 1987A. The profiles of the H$\alpha$ F656N (dashed) and the [N II] F 658N (solid) filters from the WFPC2 are shown with the filter profile of the H$\alpha$ F658N (dashed dot) from the ACS, and the H$\alpha$ and the [N II] $\lambda\lambda$ 6548 6583 doublet from our SN1987A FORS1/VLT spectrum. The emission line fluxes have been normalized to fit them in the same plot with the throughput values of the three filters.}
\end{figure}

\begin{figure*}[ht]
\includegraphics[width=9cm]{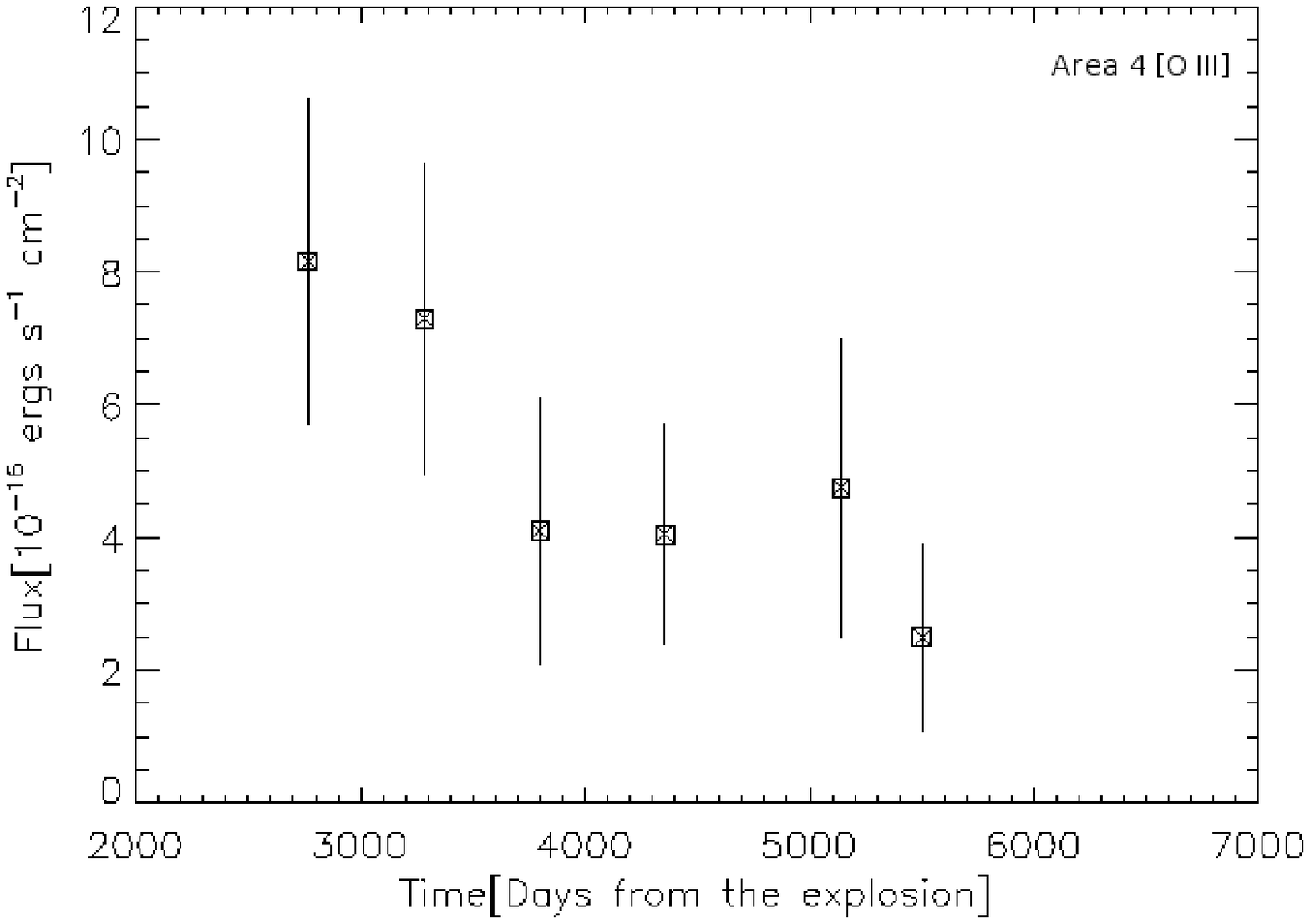}\hspace*{1mm}\includegraphics[width=9cm]{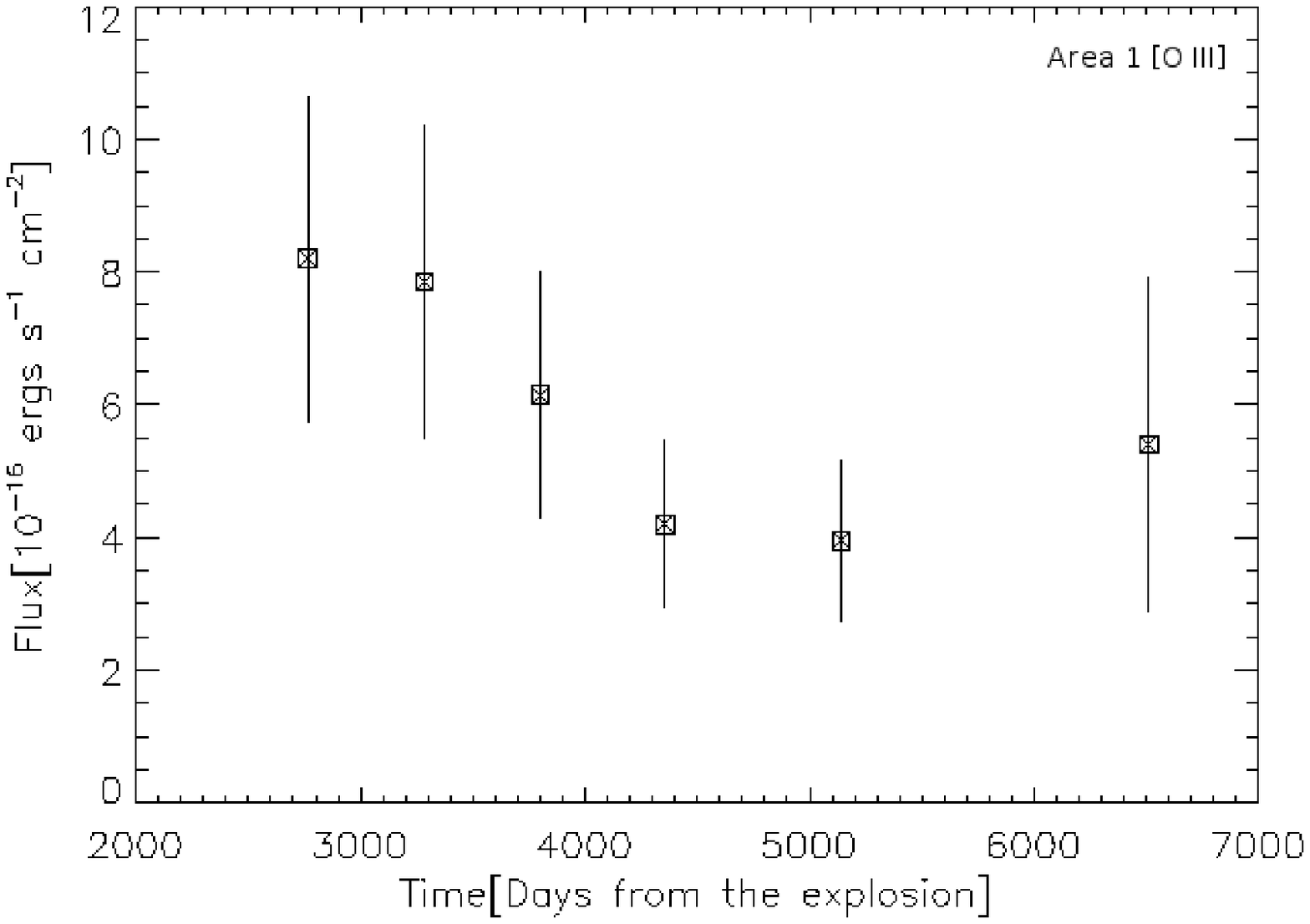}
\includegraphics[width=9cm]{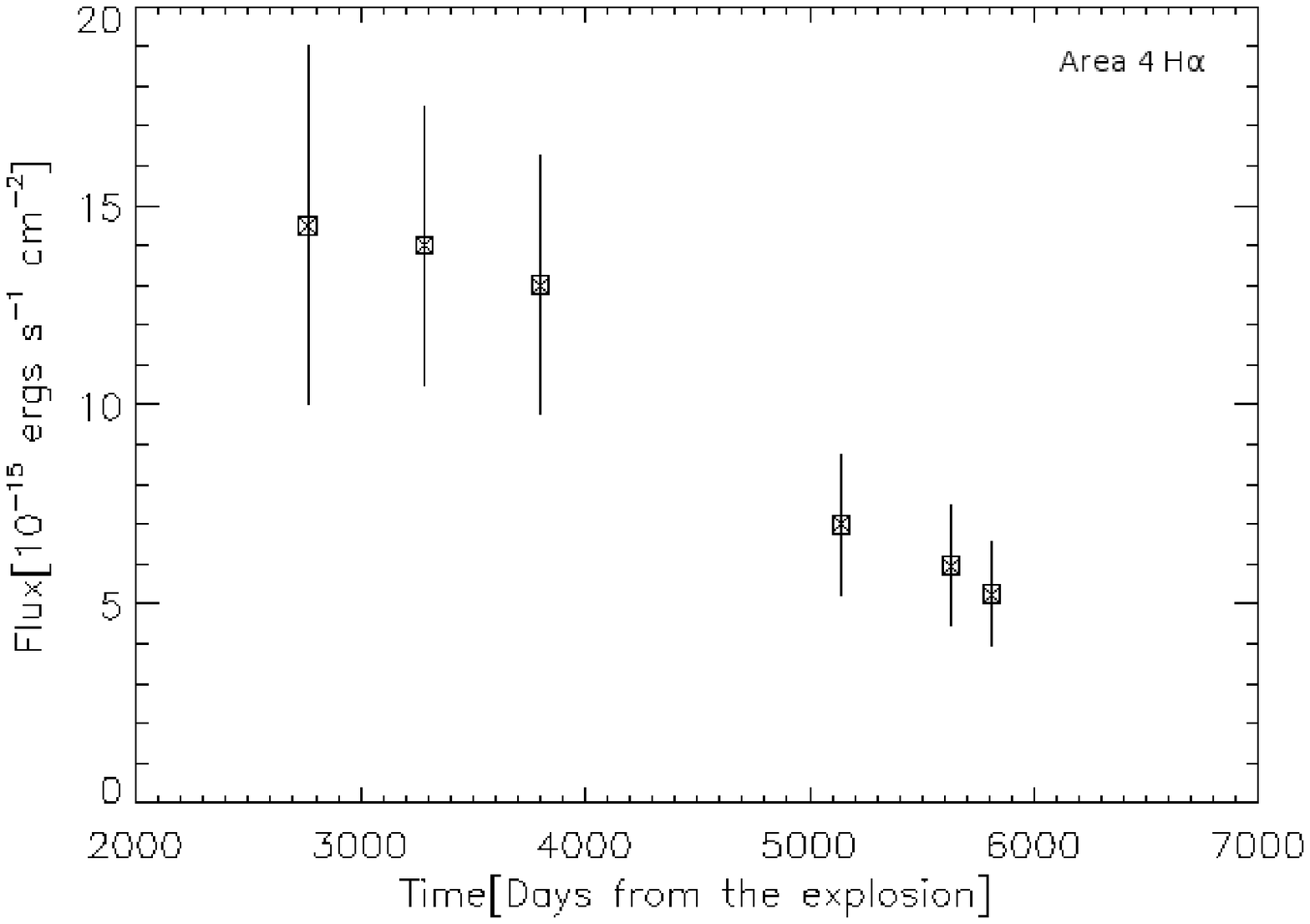}\hspace*{1mm}\includegraphics[width=9cm]{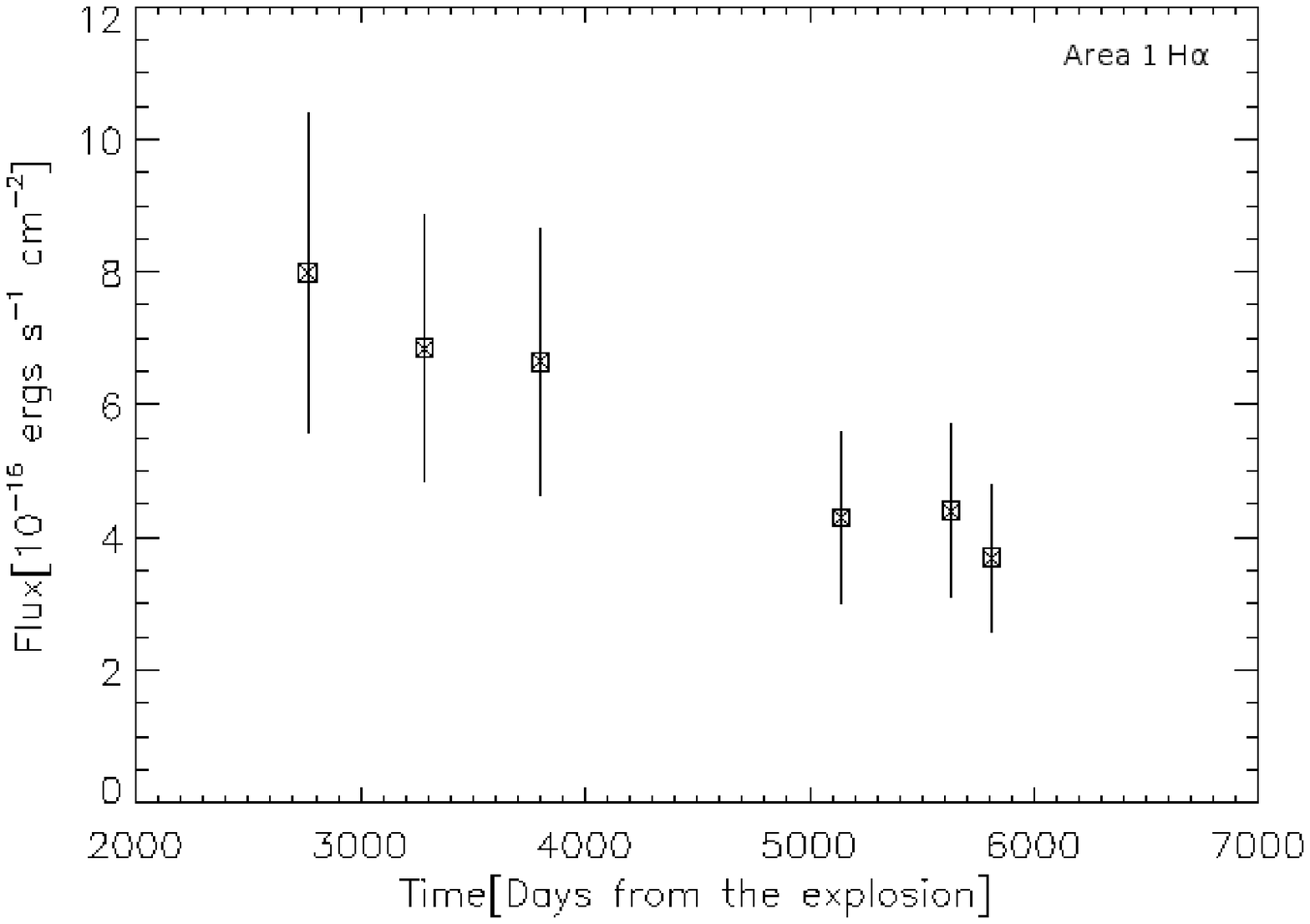}
\includegraphics[width=9cm]{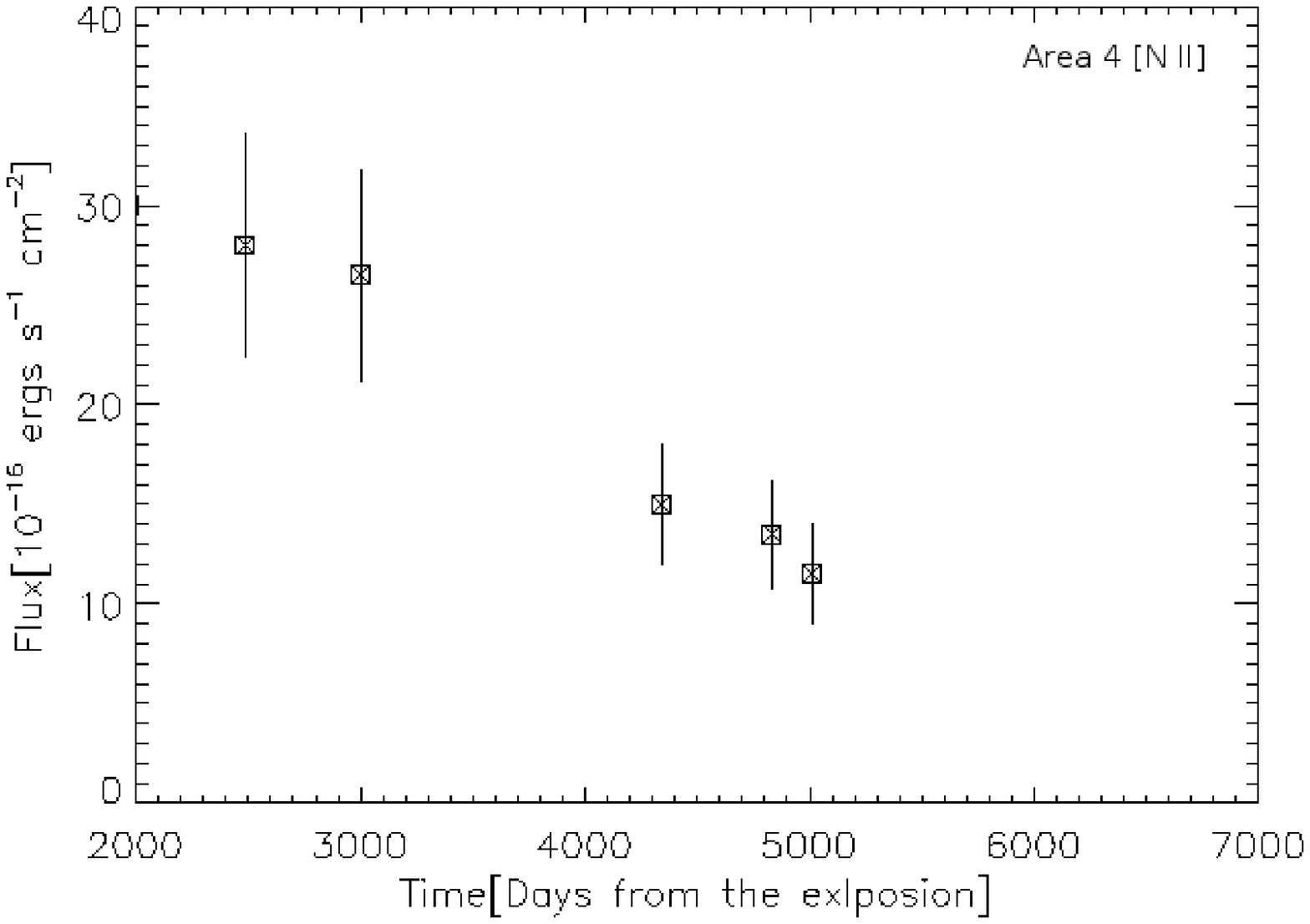}\hspace*{1mm}\includegraphics[width=9cm]{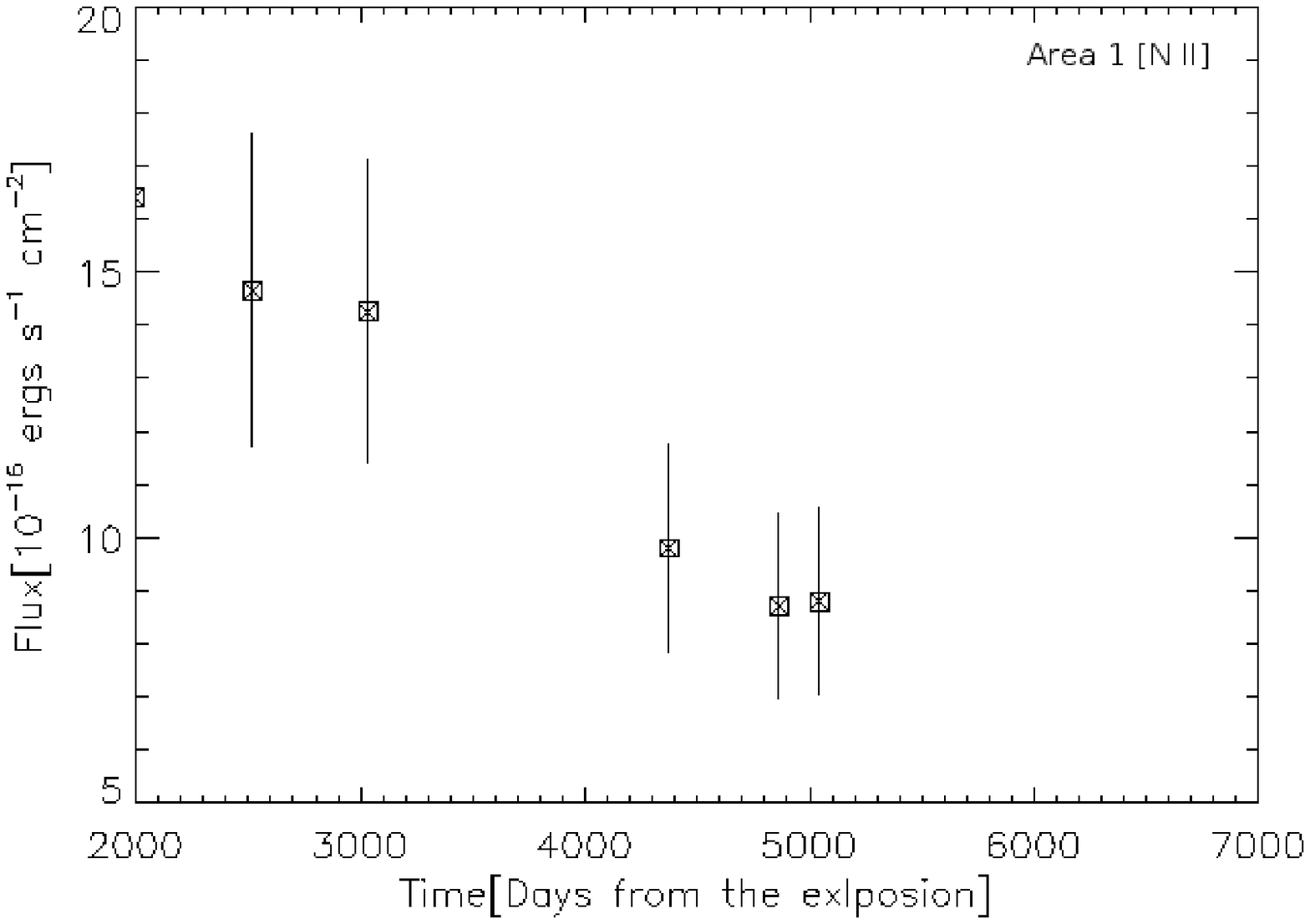}
\caption{{\it Upper panel:} [O III] light curves from the outer rings of SN 1987A. The points correspond to both the $\lambda$$\lambda$4959, 5007 \AA{} lines. {\it Middle panel:} H$\alpha$ light curves from the outer rings of SN1987A. {\it Lower panel:} [N II] light curves from the outer rings of SN~1987A. The points correspond to the $\lambda$$\lambda$6548, 6583 \AA{} line doublet. In all three panels the light curves from the southern outer ring are shown (area 4) on the left side, while on the right panels the light curves from the northern ring (area 1) are shown. Times are corrected for light travel time (see text).}
\end{figure*}

\begin{table*}
\caption{Photometric flux evolution of the outer rings of SN1987A.}
\begin{tabular}{c c c c c c c}
\hline\hline
Date & Instrument & Filter & Area 1 & Area 2 & Area 3 & Area 4$^{\mathrm{a}}$\\
\hline
940924 & WFPC2 & F502N & 6.15(1.55)$^{\mathrm{b}}$   & 8.00(2.00)& 8.75(2.20) & 6.15(1.55)$^{\mathrm{c}}$\\
1996-02-06 & WFPC2 & F502N & 5.90(1.20)  & 5.45(1.35)  & 6.30(1.60)  & 5.45(1.35)$^{\mathrm{d}}$\\
1997-07-12 & WFPC2 & F502N & 4.60(1.15)  & 3.60(0.90)  & 5.35(1.35)  & 3.50(0.90)\\
2000-06-16 & WFPC2 & F502N & 3.15(0.80)  & 3.25(0.80)  & 3.75(0.90)  & 3.05(0.75)\\
2003-01-05 & ACS   & F502N & 3.95(1.20)  &    N/A      &    N/A      & 4.75(1.20)\\
2003-11-28 & ACS   & F502N & 8.50(2.20)  & 9.60(2.45)  & 9.75(2.45)  & 2.50(0.60)\\
2004-12-15 & ACS   & F502N & 5.40(1.35)  & 9.75(2.40)  & 14.00(4.20) &    N/A    \\
1994-02-03 & WFPC2 & F656N & 8.60(2.50)  & 20.00(6.00) & 14.50(4.35) & 22.00(6.60)\\
1997-07-10 & WFPC2 & F656N & 10.50(2.10) & 16.00(3.20) & 11.00(2.20) & 19.50(3.90)\\
1998-02-05 & WFPC2 & F656N & 8.35(2.10)  & 15.00(3.00) & 11.50(2.50) & 19.00(3.80)\\
1999-01-07 & WFPC2 & F656N & 9.20(1.85)  & 15.50(3.90) & 11.00(2.75) & 17.00(4.70)\\
1999-04-21 & WFPC2 & F656N & 9.30(2.00)  & 16.50(3.30) & 10.00(2.00) & 17.50(3.50)\\
2000-01-14 & WFPC2 & F656N & 7.90(1.60)  & 13.00(2.60) & 10.50(2.10) & 13.00(2.60)\\
2000-02-02 & WFPC2 & F656N & 8.45(2.10)  & 12.50(2.50) & 11.00(2.80) & 15.50(3.90)\\
2000-06-16 & WFPC2 & F656N & 6.75(1.35)  & 12.00(2.40) & 6.25(2.00)  & 14.00(2.80)\\
2001-03-23 & WFPC2 & F656N & 8.30(2.15)  & 13.00(3.25) & 9.30(2.30)  & 13.00(2.60)\\
2001-12-07 & WFPC2 & F656N & 7.90(1.50)  & 9.75(1.95)  & 7.50(2.00)  & 11.00(2.20)\\
2002-05-10 & WFPC2 & F656N & 5.70(1.20)  & 13.50(3.30) & 9.00(1.80)  & 13.50(3.35)\\
1994-09-24 & WFPC2 & F658N & 14.50(2.90) & 21.15(4.20) & 17.50(3.50) & 26.50(5.30)\\
1996-02-06 & WFPC2 & F658N & 11.50(2.30) & 18.00(3.60) & 15.00(3.00) & 23.50(4.70)\\
1996-09-01 & WFPC2 & F658N & 11.00(2.20) & 18.50(3.70) & 15.00(3.00) & 21.00(4.20)\\
2002-05-10 & WFPC2 & F658N & 7.85(1.60)  & 12.00(2.40) & 11.00(2.20) & 12.00(2.40)\\
2003-01-05 & ACS   & F658N & 9.15(1.85)  & 15.20(3.05) & 13.50(2.70) & 14.00(2.80)\\
2003-08-12 & ACS   & F658N & 13.00(1.85) & 19.50(3.90) & 17.60(3.50) & 17.50(3.50)\\
2003-11-28 & ACS   & F658N & 12.00(2.40) & 19.50(3.90) & 17.20(3.45) & 17.00(3.40)\\
2004-12-15 & ACS   & F658N & 11.50(2.30) & 18.00(3.60) & 17.00(3.40) & 15.00(3.00)\\
2005-04-02 & ACS   & F658N & 13.00(3.25) & 17.50(3.50) & 15.50(3.90) & 17.00(4.25)\\
\hline
\end{tabular}
\begin{list}{}{}
\item[$^{\mathrm{a}}$] Flux is in units of $10^{-16}$~ergs s$^{-1}$ cm$^{-2}$
\item[$^{\mathrm{b}}$] The area corresponds to the area that the spectroscopic slit covered in the northern outer ring.
\item[$^{\mathrm{c}}$] The area corresponds to the area that the spectroscopic slit covered in the southern outer ring.
\item[$^{\mathrm{d}}$] The extinction correction was done in the same was as for the spectroscopy.
\end{list}
\end{table*}

\section{Discussion}

The emission from the ORs of SN 1987A is low compared to that from the ER (5-15\%). This of course makes the uncertainties of the measured fluxes larger, thus affecting the estimates on density and temperature. We used three independent data sets to cross-check these to see what can be used to make an analysis. A major problem for the ground-based data is the contamination from the ER, so to compile light curves of individual emission lines, HST photometry is superior. For the spectroscopy, the VLT/UVES data are particularly well suited for velocity studies, while the shorter exposures with VLT/FORS give better overall relative fluxing of spectra at a given epoch. 

As can be seen from Tables 6 and 7, the smallest errors for the velocities are for the strong red [N~II] lines and H$\alpha$. Because we know that contamination from the inner ring is worse in the blue for VLT/UVES, we rely on these lines in the red. A weighed mean for the NOR point is +292.9 km s$^{-1}$ in 2002 and +292.8 km s$^{-1}$ in 2009 with respect to the Sun. For the SOR, the corresponding numbers are +290.8 km s$^{-1}$ and +290.7 km s$^{-1}$, respectively. The NOR point (area 1, or T2 in Fig. 5) is thus $\sim 2.1$ km s$^{-1}$ redshifted compared to the SOR point (area 4, or T1 in Fig. 5). 

The centre of the supernova-ring system is at +286.6 km s$^{-1}$ according to Gr\"oningsson et al. (2008a) and +289 km s$^{-1}$ according to Crotts \& Heathcote (2000). If we use Gr\"oningsson et al. (2008a) as reference, because it partly uses the same data as we do, both areas 1 and 4 are redshifted compared to the supernova by roughly $\sim 6$ and $\sim 4$ km s$^{-1}$, respectively. The redshift of the NOR point is easy to understand from Fig. 5, whereas the redshift of area 4 is more difficult to accommodate within the geometrical model, which for the areas 1 and 4 predict $\sim 4$ and $\sim -2$ km s$^{-1}$, respectively. The difference of $\sim 6 \kms$ in Doppler shift between the model and measurement for area 4 cannot be accounted for even if we use the centre of mass velocity for SN 1987A from Crotts \& Heathcote (2000). The conclusion is that the SOR at the position of area 4 must extend to the "red" side of the ring system by $\sim 0.2-0.3$ light years rather than $\sim 0.2$ light years to the "blue" side. This increases the light-echo delay time somewhat from $\sim 620$ days to $\sim 800$ days. We used the latter value for the light delay correction in Table 8.

The densities we derive from the ground-based data, and those compiled from previous investigations show that the electron density in the ORs is likely to be $\lsim 3\times10^3$ cm$^{-3}$ (as probed by [S~II]), although the highest density, as derived from the H$\alpha$ decay, could be of the order of $\sim 5\times10^3$ cm$^{-3}$. The lowest density gas can have a density below $\sim 10^3$ cm$^{-3}$. There appears to be no obvious difference in density between the NOR and SOR, except for the gas components with the highest densities, which may have higher density in the SOR than in the NOR. The range of densities of atoms and ions in the ER is $10^3 - 4\times10^4$ cm$^{-3}$ (Lundqvist \& Fransson 1996, Mattila et al. 2010). The inner radius of the ER is $\sim 6.2\times10^{17}$ cm (cf. Lundqvist 1999). If we assume ballistic expansion of the ER out to a distance of the ORs, i.e., $\sim 1.9\times10^{18}$ cm (valid for areas 1 and 4), the atom plus ion density would be $\lsim 1.4\times10^3$ cm$^{-3}$. The electron density should be similar, which means that ballistic expansion of the ER would produce gas with a density that is at the low end of what we estimate for the ORs, especially if we need a density of $\sim 5\times10^3 \cm3$ to explain the decay of H$\alpha$ in Fig. 10. In the model of Blondin \& Lundqvist (1993) the ER is compressed as a result of interacting winds, which would indicate that a similar, or perhaps even stronger compression could have occurred as well for the ORs.

With the density of the ORs we find, the question that arises is what will happen when the supernova blast wave reaches them. In particular, will we see the same spectacular event as the optical re-brightening of the ER starting with the so-called Spot 1 in 1996 (Garnavich et al. 1997). For this to occur, transmitted shocks into the ORs must be radiative. Using Eq. 2 in Gr\"oningsson et al. (2006), we estimate the cooling time of the shocked gas to be $t_{\rm cool} \approx 83 (n_{\rm ORs}/10^3)^{-1} (V_{\rm s} / 300)^{3.4}$ years, where $n_{\rm ORs}$ is in cm$^{-3}$ and $V_{\rm s}$ is in km s$^{-1}$. If we adopt an upper limit of $n_{\rm ORs} \approx 5\times10^3$ cm$^{-3}$, $t_{\rm cool} \approx 17 (V_{\rm s} / 300)^{3.4}$ years for the densest parts of the ORs. 

In the equatorial plane the supernova blast wave expands with a velocity of $\sim 3000$ km s$^{-1}$ (cf. Gr\"oningsson et al. 2006 and references therein) once it has entered the H II region interior to the ER. The density in the H II region, $n_{\rm HII}$, is presumably lower in the direction from centre of explosion towards the ORs (cf. Mattila et al. 2010; Zhekov et al. 2010). It is therefore likely that the blast wave will move faster in those directions. Once it hits the ORs, the velocity of the transmitted shocks will simply scale as $V_{\rm blast} (n_{\rm ORs}/n_{\rm HII})^{-1/2}$. Adopting the conservative low limit for the blast wave velocity of $3000$ km s$^{-1}$ and an H II region density of 50 cm$^{-3}$, head-on transmitted shocks will have a velocity of $\sim 400$ km s$^{-1}$, or higher. Such shocks will probably not have time to be radiative before the ORs have been run over, and soft X-ray emission should offer a better chance of detecting the collision. 

For a current radius of the blast wave in the direction of the ORs of $\sim 10^{18}$ cm, the collision with the ORs in areas 1 and 4 would occur in $\sim 95~(3\times10^3/V_{\rm blast})$ years. There is, however, considerable uncertainty in the mass distribution of the circumstellar matter in the direction towards the ORs, so a collision with the ORs may occur much earlier than in $\sim 100$ years, and may only be $\lsim 20$ years away, especially in those parts of the ORs which are closest to the supernova. According to Sect. 4.1 this is likely to be the southeastern part of the NOR. Furthermore, the supernova ejecta may also expand faster in the eastern direction because the ER/ejecta interaction started there (although a slight east-west asymmetry of the ER is not ruled out). Two-dimensional hydrodynamical simulations should be made for various density distributions to investigate the time of impact and shock physics of the supernova/OR interaction.

Continued monitoring of the ring system of SN 1987A is important to see how the ejecta plough out through the circumstellar matter. While the ER is now fully dominated by the ejecta/ring interaction, the ORs are still unaffected by this, and may remain so until ejecta impact. What is  particularly interesting is to see if there is any physical connection between the three rings, or if they are truly physically separated. Some low-density gas appears to be present in the vicinity of the ER (cf. Mattila et al. 2010 and Zheov et 2010), but the distribution out to the ORs is unknown. Instruments like JWST and ALMA may not be around when the impact occurs, but their good sensitivity and spatial resolution, together with X-ray instruments like Chandra and XMM-Newton, as well as radio telescopes like ATCA, will be useful to prepare for observations when the ORs are overtaken by the ejecta of SN 1987A.  

\section{Conclusions}

We presented spectroscopic and photometric data of the ORs of SN 1987A. We extracted low and high resolution spectra of the ORs and made estimates of temperature and electron density with plasma diagnostics. Using the HST science archive data, we monitored the flux evolution of the ORs using [O~III], H$\alpha$, and [N~II] narrow band filters, and presented our results in light curve format. The temperatures and electron densities have largely remained constant for $\sim 15$ years, with the highest density being $\lsim 3\times10^3$ cm$^{-3}$, as derived from [S~II]. Additionally, the measurements of the Balmer line ratios point to recombination emission as the main power source in the ORs of the hydrogen line emission, except for H$\alpha$, which, like in the ER, also has a contribution from collisional excitation. The relatively fast decline of H$\alpha$ may require a density of $\sim 5\times10^3$ cm$^{-3}$. We find no clear difference in density and temperature for the two ORs, although the highest density in the SOR may be higher than the highest in the NOR. For an assumed distance of 50 kpc to the supernova, our geometrical model, which is based on ballistic trajectories of the outer-ring material, predicts that the distance between the supernova and the closest parts of the ORs is $\sim 1.7\times10^{18}$ cm. Those parts of the ORs probably lie in the southeastern parts of the NOR. Interaction between the supernova ejecta and these portions of the NOR could start in less than $\sim 20$ years, but may also occur considerably later. The ejecta/OR interaction will most likely be better observed in soft X-rays than in the optical.

\begin{acknowledgements}
The authors would like to thank the anonymous referee for comments that improved the manuscript, and Romano Corradi from the Isaac Newton Group of telescopes (ING) for useful comments on the data analysis of the spectra. HST data for this programme were obtained through the SINS/SAINTS programme (Robert Kirshner, PI) which have been generously supported over many cycles by NASA through grants from the Space Telescope Science Institute, which is operated by the Association of Universities for Research in Astronomy, Inc., under NASA contract NAS5-26555. The research of PL is sponsored by the Swedish Research Council.
\end{acknowledgements}

\end{document}